# Machine Learning based optimization for interval uncertainty propagation

Alice Cicirello[*], Filippo Giunta

Section of Mechanics and Physics of Structures, Faculty of Civil Engineering and Geosciences,

Delft University of Technology, Stevinweg 1, Delft 2628, NL

**Abstract**

Two non-intrusive uncertainty propagation approaches are proposed for the performance analysis of engineering systems described by expensive-to-evaluate deterministic computer models with parameters defined as interval variables. These approaches employ a machine learning based optimization strategy, the so-called Bayesian optimization, for evaluating the upper and lower bounds of a generic response variable over the set of possible responses obtained when each interval variable varies independently over its range. The lack of knowledge caused by not evaluating the response function for all the possible combinations of the interval variables is accounted for by developing a probabilistic description of the response variable itself by using a Gaussian Process regression model. An iterative procedure is developed for selecting a small number of simulations to be evaluated for updating this statistical model by using well-established acquisition functions and to assess the response bounds. In both approaches, an initial training dataset is defined. While one approach builds iteratively two distinct training datasets for evaluating separately the upper and lower bounds of the response variable, the other one builds iteratively a single training dataset. Consequently, the two approaches will produce different bound estimates at each iteration. The upper and lower response bounds are expressed as point estimates obtained from the mean function of the posterior distribution. Moreover, a confidence interval on each estimate is provided for effectively communicating to engineers when these estimates are obtained at a combination of the interval variables for which no deterministic simulation has been run. Finally, two metrics are proposed to define conditions for assessing if the predicted bound estimates can be considered satisfactory. The applicability of these two approaches is illustrated with two numerical applications, one focusing on vibration and the other on vibro-acoustics.

[*] Corresponding author.

Email: a.cicirello@tudelft.nl

## 1. Introduction

It is often the case in engineering problems that little information is available concerning the actual value and/or the inherent variability of the key input parameters of an expensive-to-run model required for a system performance assessment with respect to safety, quality, design or cost constraints, especially when the performance is assessed in terms of noise and vibration [1,2]. While the epistemic uncertainty is related to a lack of knowledge and can be potentially reduced by gathering more information, the aleatoric uncertainty is irreducible since it is related to an inherent variability of the input parameters. The performance assessment of such designs should ideally consider these uncertainties to select robust design solutions [1-2]. This can be achieved by performing the so-called forward uncertainty quantification analysis [1]. Within this analysis, the underlying uncertainties (for example, uncertainties in material properties, loading conditions, boundary conditions, and fabrication details) of the input parameters are described with an uncertainty model, and the goal is to obtain the corresponding uncertainty description of the response variables. In particular, depending on the available information on the input parameters, they can be described either with a probabilistic [1,3], non-probabilistic [4,5] (by setting upper and lower bounds [6], using an admissible convex region [7] or by means of fuzzy numbers [8-10]), or based on info-gap decision theory [11]. Alternatively, non-parametric uncertainty models can be implemented to obviate the need for a detailed description of the underlying uncertain model parameters by using some form of Random Matrix theory [1,12-14].

Broadly speaking, the forward uncertainty propagation approaches can be grouped into intrusive and non-intrusive approaches. Non-intrusive approaches do not require direct modifications on the expensive-to-evaluate model which can be treated as a black-box function. The model is evaluated for each realization of the uncertainty variable, to assess the input uncertainty effects on the response variables. One very well-known example when dealing with probabilistic uncertainty descriptions is the Monte Carlo sampling strategy [3] that can be used to yield the probability density function and other statistics of a response variable of interest. On the other hand, the intrusive approaches require modifications of the governing equations of the problem to build models that inherently embed certain uncertainty descriptions. Very well-known examples are some of the expansion-based methods [15]. These approaches enable a drastic reduction of the computational cost of the uncertainty propagation and are valid under certain assumptions on the uncertainty levels.

Although many approaches have been developed in the last decades, there are still unresolved challenges that need to be tackled for assessing the noise and vibration performance of an engineering design, especially when uncertain input parameters are described by interval variables [5]. Intervals are used for describing parameters not known precisely for which only the maximum and minimum values, also known as bounds, they could take is known. These bounds are often representing "tolerance values" on the parameters associated with the properties of a component of a system. Bounds can also be used to describe known limits on the values that an input variable of a model might take. This type of bounded description is commonly used in engineering practices, especially at early design stages when a limited amount of data and/or information may be available to select the form of the distribution of the uncertain variable and/or to specify the parameters of the distribution. In fact, the results on a system performance assessment with respect to safety, quality, design or cost constraints are strongly dependent on the probability density function assumed [16,17]. Often it is assumed that the interval description is equivalent to the uniform distribution, by exploiting the Maximum Entropy Principle [18] so that well-established probabilistic uncertainty quantification approaches can be implemented. However, converting an interval description into a uniform probability distribution means assigning equal probability of occurrence to each value within the interval (see for example [19]). Moreover, if the upper bound of the response is obtained at the upper bound of an interval variable, there is a very low probability of the random sampling strategy selecting this value precisely [19].

This paper is focused on engineering problems characterized by an expensive-to-run deterministic model for which the input parameters are described by interval variables to yield the bounds on a response variable of interest. No assumption on the monotonicity of the response with respect to the input uncertainties is made, since this is not often met when dealing with noise and vibration problems because of the presence of resonances [19]. Furthermore, the attention is focused on non-intrusive interval uncertainty propagation strategies. Among the state-of-the-art approaches, the Interval Vertex Method (VM) [20] would not provide an accurate estimate of the response bounds because of the aforementioned non-monotonic behaviour. This is because, as the name suggests, the VM requires evaluating the deterministic model only for the combination of all the vertices of the interval input variables. Alternatively, the Subinterval Method (SM) [19] can be used. With this approach, the domain of each interval variable is decomposed into subintervals within which it is assumed that the generic quantity of interest is a monotonic function with respect to the uncertain parameters. Then, the response bounds are obtained as the minimum and maximum across all the possible combinations of the endpoints of the subintervals. Since it is not known a priori for which values of the interval variables the bounds of the generic response quantity of interest are attained, then enough subintervals should be considered. Global Search Algorithms [5] can be also employed to deal with these problems. These algorithms are usually combined with a surrogate model which approximate the relationship between the structural response and the input parameters and is computationally cheaper to run [5,21]. Such surrogate models can be obtained by using high-order polynomial function or are based on machine learning techniques [5,21]. For example, Radial Basis Function [22], Kriging interpolation schemes [23] and Artificial Neural Network [24,25] have been implemented to address the interval uncertainty propagation problem. Nonetheless, these approaches would require to evaluate a large number of runs to construct an accurate surrogate model, and to fine-tune the surrogate model parameters. Therefore, they can entail a considerable computational effort if a large number of uncertain parameters need to be considered. De Munck et al. [26] proposed an adaptive response surface method based on Kriging model to yield an accurate meta-model with a limited computational cost. In [27] the upper and lower bounds of a black-box response function were considered as two separate optimization problems, and the computational cost was reduced by applying the Efficient Global Optimization (EGO) algorithm [28]. Recently, a similar approach has been used to evaluate the response bounds for train-bridge systems [29].

In this paper, two non-intrusive UQ propagation approaches are proposed for the analysis of a generic system subject to interval uncertainties. These approaches employ a machine learning based optimization strategy for evaluating the Upper Bound (UB) and Lower Bound (LB) of a generic response variable over the set of the possible responses by iteratively selecting a limited number of simulations to be evaluated. The response is considered as a black-box function that depends on the interval variables for which no analytical solution or derivative information is available. No assumptions on the linearity or monotonicity of the response variable with respect to the interval uncertainties are required. It is assumed that the response function is smooth and it can be obtained for discrete values of the interval variables by running an expensive-to-evaluate deterministic model. However, in engineering practice, only a limited number of simulation evaluations can be carried out because of computational budget and/or time constraints. The proposed approaches embed the lack of knowledge caused by not evaluating the response function for all the possible combinations of the interval variables by developing a probabilistic description of the response by using the so-called Gaussian Progress (GP) regression model [30]. An iterative procedure which uses the GP and the so-called Acquisition Functions (AFs) is then used to update this statistical model to obtain accurate estimates of the response bounds. Therefore, the interval uncertainty propagation is framed into a Bayesian Optimization problem [31,32]. Although the use of GP and AFs for solving optimization problems have been investigated for a long time in the machine learning community [28, 31-34], and have recently been explored in engineering applications [26,27, 29], there are still open questions for implementing this framework for interval uncertainty quantification of engineering systems, such as: (i) How many simulations should be run initially and how should they be selected? (ii) What are the effects of employing certain acquisition functions? (iii) What is the effect of considering two separate GPs for determining the estimates on the lower and upper response bounds versus considering a single GP? (iv) What is the best way to represent the response bound estimates for effectively communicating the results to engineers? (v) How to assess if the estimated bounds are satisfactory? These questions are addressed in the present work.

The proposed approaches employ one of the Design of Experiments techniques [21,35,36], the Taguchi method, to select the simulations needed to construct the initial training dataset before any simulation results is available. This initial training dataset is used to build a GP regression model which is fully defined in terms of the exponential of a weighted distance kernel function. Three AFs (Probability of Improvement [21,28, 37], Expected Improvement [28, 34, 37] and Confidence bounds [38,39]) are explored to identify the new simulations that need to be run to improve the bounds prediction. Once the additional simulations are carried out, these results are used to build the so-called enhanced training dataset. In Approach A, two separate enhanced training datasets (one for the UB and one for the LB) are defined. Approach B uses a single enhanced training dataset. For each enhanced training dataset, the GP and AFs are used to select the new simulations to be evaluated to improve the response bounds prediction. This iterative procedure is stopped when (i) the entire computational budget available has been used; and/or (ii) a threshold on the absolute maximum value of the AF is reached. The response bound estimates are then expressed in terms of the maximum and minimum of the posterior mean function of the corresponding GP. Moreover, the confidence bounds at the points are used to quantify the remaining epistemic uncertainty. Two metrics are then introduced to assess if the response bounds obtained are satisfactory or if additional simulations need to be carried out.

The paper is structured as follows: the two proposed approaches are described in section 2. The algorithms summarizing the steps of Approach A and Approach B are described in section 3 and section 4, respectively. The two approaches are illustrated by investigating a single degree of freedom system and a plate-acoustic cavity subject to interval uncertainties in section 5. Finally, section 6 and section 7 provide discussions and conclusions, respectively.

## 2. Interval uncertainty propagation using machine learning optimization
### 2.1 Problem description

Let us consider a generic engineering system represented by a deterministic model (obtained with Finite Element [40], Boundary Element [41], Wave Finite Element [42] or any other deterministic modelling strategy) subject to time-varying inputs and characterised by $r$ independent uncertain input parameters described as intervals. These uncertain input parameters are collected into the interval vector $\mathbf{b} = \left[\underline{\mathbf{b}}, \overline{\mathbf{b}}\right] = \left[b_1^{\text{int}}, b_2^{\text{int}}, \ldots, b_r^{\text{int}}\right]^{\text{T}}$ where $\underline{\mathbf{b}}$ and $\overline{\mathbf{b}}$ denote the vectors collecting the lower bound and the upper bound of the $i$-th interval variable $b_i^{\text{int}} = \left[\underline{b}_i, \overline{b}_i\right]$, $(i = 1, 2, \ldots, r)$. The generic response quantity of interest (e.g. time or frequency domain representation of displacement, velocity, acceleration, etc.) of this system can be written as $w(\mathbf{b})$, indicating explicitly its dependence on the interval uncertain parameters $\mathbf{b}$. The Lower Bound (LB), $\underline{w}$, and Upper bound (UB), $\overline{w}$, of the so-called objective function $w(\mathbf{b})$ are obtained by solving the optimization problems:

$$\underline{w} = \min_{\underline{\mathbf{b}} \leq \mathbf{b} \leq \overline{\mathbf{b}}}\left[w(\mathbf{b})\right] \qquad \overline{w} = \max_{\underline{\mathbf{b}} \leq \mathbf{b} \leq \overline{\mathbf{b}}}\left[w(\mathbf{b})\right] \qquad (1)$$

In other words, $\underline{w}$ and $\overline{w}$ are, respectively, the minimum and the maximum values of the all possible response evaluations obtained when each uncertain parameter $b_i^{int}$, $(i=1,2,...,r)$ varies independently over its range. The direct solution of the optimization problems in Eq. (1) can become challenging to the point of being unfeasible for large complex models of engineering systems. This is because the analytical expressions of $w(\mathbf{b})$ and of its derivatives with respect to specific combinations of the interval parameters $\mathbf{b}_j$ are often not available. Consequently, the resolution of the optimization problem would require multiple runs of the underlying model or of a surrogate model [5,21]. However, as the number of uncertain parameters increases and as the mathematical model complexity increases, the evaluation of the bounds on $w(\mathbf{b})$ can become computationally expensive and prone to errors [5,21]. These problems are particularly relevant when dealing with large numerical models for the vibro-acoustics performance assessment of engineering systems. Moreover, vibro-acoustic models display a non-monotonic behaviour of the response due to the presence of resonant peaks that might occur in certain frequency bands for different combinations of the interval input parameters.

## 2.2 Interval Uncertainty Quantification as a Bayesian optimization problem

The proposed approaches frame the uncertainty quantification under interval variables into a Bayesian Optimization (BO) framework [31,32]. The BO is an approach to maximising an objective function. It is based on building a probabilistic surrogate of an objective function by using a Bayesian machine learning technique, the Gaussian Process Regression [30]. The so-called Acquisition Functions are then applied to this probabilistic model for efficiently selecting the objective function samples to be evaluated [31,32].

To transform $w(\mathbf{b})$ in Eq. (1) in a probabilistic model, the objective function $w(\mathbf{b})$ is first approximated with a discretized function which would be obtained by evaluating the response variable $w(\mathbf{b}_j)$ at several $\mathbf{b}_j$ combinations of the interval variables. No assumption is made about the linearity or monotonicity of the response variable with respect to the interval uncertainties. Since only few evaluations of $w(\mathbf{b}_j)$ might be carried out, each $w(\mathbf{b}_j)$ can be considered as uncertain until it is evaluated. Therefore, this uncertainty is not caused by noise in the observations or uncertainties related to the assumptions in the physics-based model. Instead, it embeds the lack of knowledge caused by not having evaluated $w(\mathbf{b}_j)$ for every $\mathbf{b}_j$ by running the deterministic model. Each $w(\mathbf{b}_j)$ can be then modelled as a Gaussian random variable. The collection of Gaussian random variables $w(\mathbf{b}_j)$ represents a Gaussian Process (GP) of $w(\mathbf{b})$. The GP represents a probability distribution over the function $w(\mathbf{b})$, that is the prior distribution $p(w(\mathbf{b}))$. The GP regression is a non-parametric approach that finds a distribution for $w(\mathbf{b})$ over the possible interpolating functions that are consistent with the observed data $D$ [30]. This distribution is the so-called posterior distribution and is indicated with $p(w(\mathbf{b})|D)$. Here, $D$ is obtained by running some deterministic simulations of the model for fixed $\mathbf{b}_j$. After some observations $D$ become available, the $p(w(\mathbf{b})|D)$ is obtained by means of Bayesian inference:

$$p(w(\mathbf{b})|D) = \frac{p(w(\mathbf{b}), D)}{p(D)} = \frac{p(D|w(\mathbf{b}))p(w(\mathbf{b}))}{p(D)} \tag{2}$$

where $p(w(\mathbf{b}), D)$ is the joint distribution of $w(\mathbf{b})$ and $D$, that can be expressed as the product of $p(D|w(\mathbf{b}))$, that is the so-called likelihood model, and $p(D)$, that is the marginal likelihood (or evidence). $p(D)$ is independent on $w(\mathbf{b})$ and is therefore a normalizing constant. Within the GP regression model [30], we are dealing with Gaussian distributions which are closed under conditioning. This means that the resulting predictive posterior distribution is also Gaussian, and the expression of its mean and variance are readily available [30].

The goal of interval uncertainty quantification is to obtain the maximum and the minimum of $w(\mathbf{b})$ at two distinctive combinations of the interval variables $\mathbf{b}_j$, ideally by evaluating the minimum number of simulations. Therefore, there is a need for a strategy to select adaptively a limited number of deterministic simulations of the model to be included in $D$ to yield the best estimate of the response bounds. In turn, this would result in locally improving the posterior of possible interpolating functions $p(w(\mathbf{b})|D)$ to yield an estimate of the bounds of $w(\mathbf{b})$ in terms of the maximum or minimum of the predictive posterior mean function, and to quantify the uncertainty associated with that estimate in the form of confidence bounds. As in BO, this can be achieved by using the Acquisition Functions (AFs). These ad-hoc functions of the GP regression model are inexpensive to evaluate compared to the objective function. The AFs are used to select which combination of the interval variables $\mathbf{b}_j$ has to be evaluated next in order to improve the most the assessment of each response bound.

Based on this framework, two approaches are proposed and are schematically shown in Figure 1 and Figure 2. Both approaches require the selection of an initial number of simulations to construct the so-called initial training dataset. This dataset is used to build a GP regression model of the response variable. Subsequently, the GP is used in combination with the AFs to identify two combinations of the interval variable at which additional simulations must be evaluated to improve the accuracy of the predicted response bounds. These simulations are then used to enhance the training dataset, and to identify through an iterative procedure the next set of simulations to be performed to improve the accuracy of the predicted response bounds. Stopping criteria are used to limit the number of simulations to be evaluated. The difference between these two approaches is in the use of the simulation results obtained at each iteration. Approach A considers two separate enhanced training datasets (one for the evaluation of the UB and one for the evaluation of the LB) and therefore addresses the two optimization problems in parallel. Approach B uses a single enhanced training dataset. As a result, at the end of each iteration, different predictive posterior distributions will be obtained with the two approaches which consequently will yield different bound estimates.

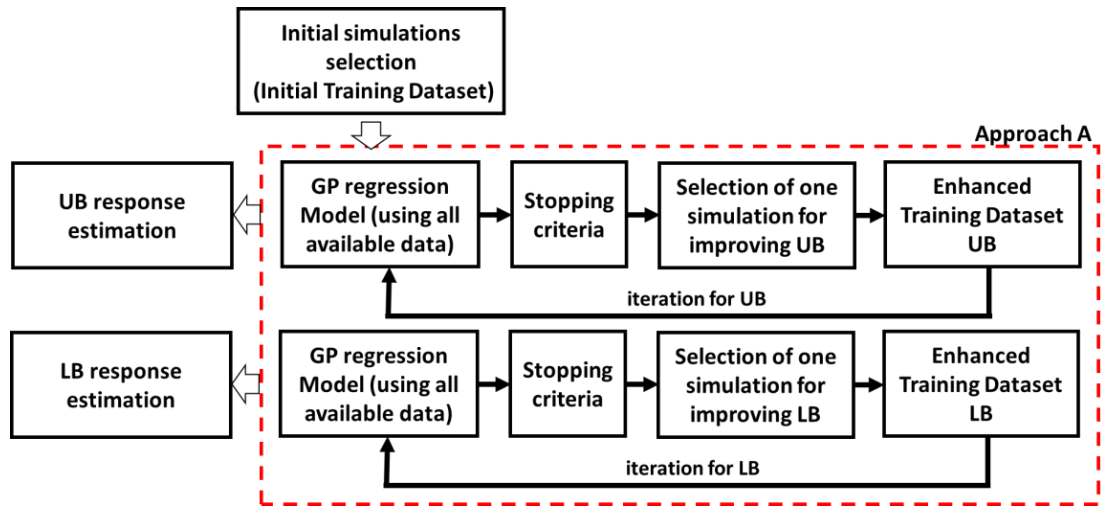

Figure 1 Schematic representation of Approach A.

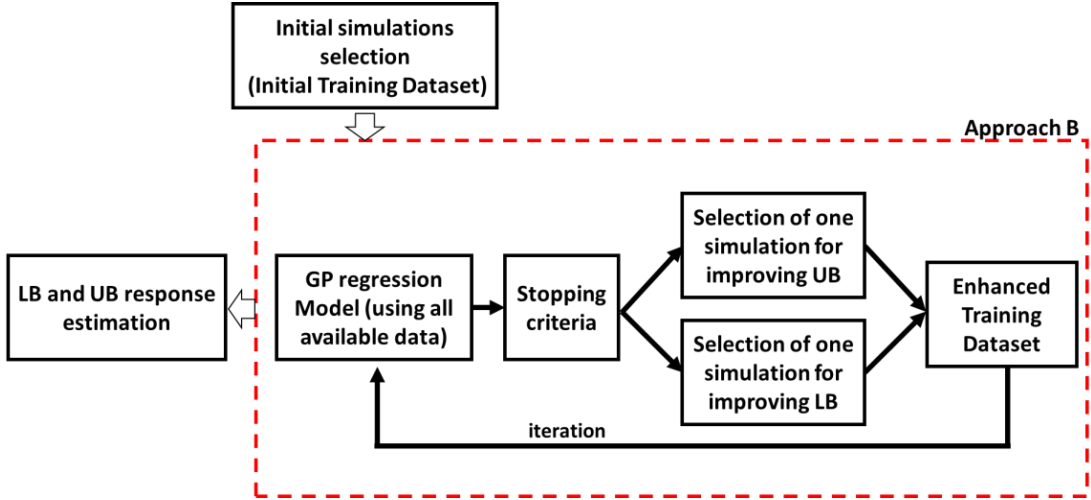

Figure 2 Schematic representation of Approach B.

Finally, two metrics are introduced to define conditions for assessing if the predicted response bounds are satisfactory or additional simulations need to be evaluated.

The two approaches share the same building blocks that are described in what follows.

## 2.3 Gaussian Process regression model

The *n* Gaussian random variables representing each $w(\mathbf{b}_j)$ are collected into a vector

$$\mathbf{w} = \left[ w(\mathbf{b}_1), w(\mathbf{b}_2), \ldots, w(\mathbf{b}_n) \right] \tag{3}$$

This collection of variables has a jointly Gaussian distribution. Therefore, the vector $\mathbf{w}$ represents a GP of the response function $w(\mathbf{b})$. This representation enables the prediction of responses $\mathbf{w}_*$ at certain values of the input interval variables $\mathbf{b}_{*j}$ for which the deterministic simulations have not been performed. In particular, without loss of generality, the vector $\mathbf{w}$ is partitioned into $\mathbf{w} = [\mathbf{w}_0, \mathbf{w}_*]$: (i) a vector $\mathbf{w}_0$ of noise-free observations of the function $w_{0j} = w(\mathbf{b}_{0j})$ obtained by evaluating $n_0$ simulations for each element of $\mathbf{B}_0 = [\mathbf{b}_{01}, \ldots, \mathbf{b}_{0n_0}]$; (ii) and a vector $\mathbf{w}_*$ of $n_*$ function realisations $w_{*k} = w(\mathbf{b}_{*k})$ that we want to predict at each element of $\mathbf{B}_* = [\mathbf{b}_{*1}, \ldots, \mathbf{b}_{0n_*}]$ without running any simulations. By using the GP regression model, the underlying distribution of $\mathbf{w}$ (joint distribution of $\mathbf{w}_0$ and $\mathbf{w}_*$) can be learned for a given a training dataset $D = \{(\mathbf{b}_{0j}, w_{0j}), j = 1 : n_0\}$ of noise-free observations, and a test dataset of function realisations $\mathbf{W}_*$ we want to predict. Because of the properties of the multivariate Gaussian distribution, the joint distribution of $\mathbf{W}_0$ and $\mathbf{W}_*$ is itself Gaussian [30], and it can be expressed as:

$$\begin{pmatrix} \mathbf{w}_0 \\ \mathbf{w}_* \end{pmatrix} \sim N\left( \begin{bmatrix} \boldsymbol{\mu}_0 \\ \boldsymbol{\mu}_* \end{bmatrix}, \begin{bmatrix} \boldsymbol{\Sigma}_{00} & \boldsymbol{\Sigma}_{0*} \\ \boldsymbol{\Sigma}_{*0} & \boldsymbol{\Sigma}_{**} \end{bmatrix} \right) \tag{4}$$

where $\boldsymbol{\mu}_0 = \left[ m_0(\mathbf{b}_{01}), \ldots, m_0(\mathbf{b}_{0n_0}) \right]$ and $\boldsymbol{\mu}_* = \left[ m_*(\mathbf{b}_{*1}), \ldots, m_*(\mathbf{b}_{*n_*}) \right]$ are the mean functions evaluated at the training dataset $D = (\mathbf{B}_0, \mathbf{w}_0)$ and test dataset $(\mathbf{B}_*, \mathbf{w}_*)$, respectively, and $\boldsymbol{\Sigma}_{0*}$ is the matrix of size $n_0 \times n_*$ of the covariance, Cov, evaluated at all pairs of training and test points

$$(\boldsymbol{\Sigma}_{0*})_{i,j} = \text{Cov}\left[ w(\mathbf{b}_{0i}), w(\mathbf{b}_{*j}) \right] = \kappa(\mathbf{b}_{0i}, \mathbf{b}_{*j}) \tag{5}$$

The other covariance matrices are defined in a similar fashion:

$$(\boldsymbol{\Sigma}_{00})_{i,j} = \text{Cov}\left[ w(\mathbf{b}_{0i}), w(\mathbf{b}_{0j}) \right] = \kappa(\mathbf{b}_{0i}, \mathbf{b}_{0j}) \tag{6}$$

$$(\boldsymbol{\Sigma}_{**})_{i,j} = \text{Cov}\left[ w(\mathbf{b}_{*i}), w(\mathbf{b}_{*j}) \right] = \kappa(\mathbf{b}_{*i}, \mathbf{b}_{*j}) \tag{7}$$

The covariance matrix defines the similarity between the *i*-th and *j*-th entries of the random variable. Therefore, it encodes the prior assumption on the smoothness of the black-box response function. The covariance matrix is defined in terms of a positive definite kernel function $\kappa$ [30, 43] which for a pair of points (e.g. two training points) return the similarity between those points as a scalar.

Because of the properties of the multivariate Gaussian distribution, it can be easily shown that the posterior predictive distribution of $\mathbf{w}_*$, given the training dataset $D = (\mathbf{B}_0, \mathbf{w}_0)$ and the chosen test points $\mathbf{B}_*$, can be obtained by conditioning [30]:

$$p(\mathbf{w}_* | \mathbf{B}_*, \mathbf{B}_0, \mathbf{w}_0) \sim N\left( \boldsymbol{\mu}_* + \boldsymbol{\Sigma}_{*0} \boldsymbol{\Sigma}_{00}^{-1} (\mathbf{w}_0 - \boldsymbol{\mu}_0), \boldsymbol{\Sigma}_{**} - \boldsymbol{\Sigma}_{*0} \boldsymbol{\Sigma}_{00}^{-1} \boldsymbol{\Sigma}_{0*} \right) \tag{8}$$

This equation takes into account the mean and the variance at each test point, as well as the correlation between any pair of test points. Only the marginal variances ($\text{diag}(\boldsymbol{\Sigma}_{**} - \boldsymbol{\Sigma}_{*0} \boldsymbol{\Sigma}_{00}^{-1} \boldsymbol{\Sigma}_{0*})$) are needed to measure of the predictive uncertainty at each test point. Therefore, the predictive marginal mean and variance for a single test point $\mathbf{b}_j$ are expressed, respectively, as:

$$m(\mathbf{b}_j) = m(\mathbf{b}_{*j}) + \mathbf{\Sigma}_{j0}\mathbf{\Sigma}_{00}^{-1}(\mathbf{w}_0 - \boldsymbol{\mu}_0) \qquad (9)$$

$$\sigma^2(\mathbf{b}_j) = \kappa(\mathbf{b}_{*j}, \mathbf{b}_{*j}) - \mathbf{\Sigma}_{j0}\mathbf{\Sigma}_{00}^{-1}\mathbf{\Sigma}_{0j} \qquad (10)$$

where $\mathbf{\Sigma}_{j0}$ is a matrix of size $1 \times n_0$. From Eqs. (9-10), it is possible to observe that while each element of the mean vector is shifted by the conditioned variable, the covariance matrix is independent on the conditioned variable. The posterior distribution obtained will be such that the GP will have zero variance and the mean corresponding to the actual response for each training dataset point.

Without loss of generality, it can be assumed that $\boldsymbol{\mu}_* = \mathbf{0}$ and $\boldsymbol{\mu}_0 = \mathbf{0}$. The kernel function is the crucial element of the GP since it determines the characteristic shape of the function being approximated [30,43]. The selection of a type of kernel function encodes the user's belief on the function being predicted [30]. Broadly speaking, kernel functions can be subdivided into stationary and non-stationary [30,43]. While stationary kernels are invariant to translation and the covariance is dependent only on the relative position of $\mathbf{b}_j, \mathbf{b}_\ell$, this is not the case for non-stationary kernel. The simplest non-stationary kernel is the linear kernel [30,43]. In this paper, the kernel chosen is stationary and is defined as the exponential of a weighted distance function (also known as squared exponential), which is preferred when little is known about the black-box function to be approximated [21,28]:

$$\kappa(\mathbf{b}_j, \mathbf{b}_\ell) = \exp\left[-d(\mathbf{b}_j, \mathbf{b}_\ell)\right], \qquad (11)$$

where the weighted distance function is defined as [21, 28]:

$$d(\mathbf{b}_j, \mathbf{b}_\ell) = \sum_{h=1}^{r} \theta_h \left|b_{j,h} - b_{\ell,h}\right|^{p_h}, \quad (\theta_h \geq 0 \text{ and } p_h \in [1,2]) \qquad (12)$$

which depends on two sets of unknown hyperparameters collected into the $r$-vectors $\boldsymbol{\theta} = \begin{bmatrix} \theta_1 & \theta_2 & \ldots & \theta_r \end{bmatrix}^T$ and $\mathbf{p} = \begin{bmatrix} p_1 & p_2 & \ldots & p_r \end{bmatrix}^T$. In Eq. (12) $b_{j,h}$ and $b_{\ell,h}$ indicate the $h$th component of the vectors $\mathbf{b}_j$ and $\mathbf{b}_\ell$, while $\theta_h$ and $p_h$ are respectively the $h$th component the vectors $\boldsymbol{\theta}$ and $\mathbf{p}$. Essentially, the hyperparameters $\theta_h$ describe how sensitive is the quantity of interest with respect to each variable $b_{j,h}$ [21,28]. The hyperparameter $p_h$ rules the smoothness of the correlation. The higher the value of $p_h$, the smoother the correlation. A detailed explanation of the role played by the hyperparameters $\theta_h$ and $p_h$, can be found in references [21,28].

These hyperparameters $\boldsymbol{\alpha} = [\boldsymbol{\theta}, \mathbf{p}]$ can be estimated by maximising the log marginal likelihood (or equivalently, minimising the negative log marginal likelihood) of the observed data [30,43,44]. Independently of the kernel chosen, the log marginal likelihood of a Gaussian distribution and its partial derivatives with respect to the hyperparameters can be expressed in closed-form as [30]:

$$L(\boldsymbol{\alpha}) = -\frac{1}{2}\mathbf{w}_0^T \boldsymbol{\Sigma}_{00}^{-1} \mathbf{w}_0 - \frac{1}{2}\ln|\boldsymbol{\Sigma}_{00}| - \frac{n_0}{2}\ln(2\pi) \tag{13}$$

$$\frac{\partial}{\partial \alpha_j} L(\boldsymbol{\alpha}) = \frac{1}{2}\mathbf{w}_0^T \boldsymbol{\Sigma}_{00}^{-1} \frac{\partial \boldsymbol{\Sigma}_{00}}{\partial \alpha_j} \boldsymbol{\Sigma}_{00}^{-1} \mathbf{w}_0 - \frac{1}{2}\mathrm{Tr}\left(\boldsymbol{\Sigma}_{00}^{-1} \frac{\partial \boldsymbol{\Sigma}_{00}}{\partial \alpha_j}\right) \tag{14}$$

where $|\bullet|$ indicates the determinant, while $\mathrm{Tr}(\bullet)$ is the trace operator. Setting to zero the closed-form expression of the partial derivatives, it is then possible to evaluate numerically the estimates of $\boldsymbol{\alpha}$. It is well known that the log marginal likelihood is a non-convex optimization problem can be characterised by several local optima [31-32, 43-45]. The selection of a point different from the global optimum can considerably affect the accuracy of the hyperparameters, and consequently of the GP regression model. Therefore, it is recommended to consider multi-starting points optimization strategies to evaluate the global optimum [43,44]. Alternative approaches for hyperparameters estimation include maximum a-posteriori [43] and stochastic-based inference methods [45,46].

While all the quantities of interests for building the GP are defined via closed-form expressions, their computation can become very challenging for large training dataset due to the $O(n_0^3)$ time complexity of the determinant and inverse of the covariance matrix of the training data [30]. Reducing the computational cost while preserving the accuracy of the GP is an active area of research in computer science [47,48]. Since the response variable of a large-scale-model might be high-dimensional (meaning that it is defined as a function of a large number of interval variables), the size of both training and test datasets could be quite large. Therefore, the complexity in each optimization step could hinder the application of the GP regression model. In this paper, this problem is addressed by selecting the minimum number of elements required for the initial training dataset, and subsequently for the enhanced training dataset, to improve the GP capability to predict the response bounds while containing the computational costs, as described in what follows.

**2.4 Initial training dataset**

The training data $D = (\mathbf{B}_0, \mathbf{w}_0)$ is the set of observed results obtained by evaluating the deterministic model $\mathbf{w}_0$ for specific combinations of the interval parameters $\mathbf{B}_0$. When only one interval variable is considered, the interval can be simply subdivided into two subintervals of the same size, resulting into three simulations to be evaluated. When many interval variables are considered, selecting the effective combinations of the interval parameters before any runs of the underlying mathematical model are performed is more challenging. In general, the generic response quantity of interest is a non-linear and non-monotonic function with respect to the interval parameters. Therefore, the effects of one interval parameter on the generic response quantity of interest depend on the values taken by the other parameters. In other words, there can be interactions between the uncertain parameters. Although advanced Design of Experiments techniques [21,35,36] can handle these interactions and select effectively the minimum number of interval variables combinations, and thus the corresponding simulations to be evaluated, in this paper it is assumed that no prior information about the interactions between the interval parameters and their effect on the response is available. These interactions are then "learned" during the iterative steps of the proposed approaches. The Taguchi method [35,36] is implemented here to (i) cover as much as possible the *r*-dimensional hypercube represented by *r>2* interval variables and (ii) to limit the initial computational effort required for building a generic initial GP regression model of the response function.

The Taguchi method [35,36] is a well-established Design of Experiment technique which selects the so-called design points, that are the simulations that are going to be run, by using the properties of the so-called orthogonal arrays. This approach requires specifying a number of *q* levels, which would result in *q-1* subintervals within each interval variable. The minimum number of design points *s* (neglecting the interaction between the interval parameters) corresponds to $s = 1 + r(q-1)$.

However, it is desired that the Taguchi matrix $\mathbf{L}_s(q^r)$ is balanced, which results in $s$ being a multiple of $q$. Interested readers can find more details about the Taguchi method and approaches for deriving the Taguchi matrix in reference [35,36]. The selection on the number of levels $q$, and the resulting $s$ simulations, is strongly dependent on the total number of simulations available for the problem under investigation. As shown in the second numerical application, for a limited total number of simulations, increasing the number of $q$ does not necessarily improve the bounds prediction. As a rule-of-thumb, it is recommended to not use more than a quarter of the total number of simulations available for the initial training dataset.

The set of $s$ combination of the interval variables $\mathbf{b}_j$ are collected in the matrix $\mathbf{L}_s(q^r)$ of dimension $s \times r$

$$\mathbf{L}_s(q^r) = \begin{bmatrix} b_{1,1} & b_{1,2} & \ldots & b_{1,r} \\ b_{2,1} & b_{2,2} & \ldots & b_{2,r} \\ \vdots & \vdots & \vdots & \vdots \\ b_{s,1} & b_{s,2} & \ldots & b_{s,r} \end{bmatrix} = \begin{bmatrix} \mathbf{b}_{01} \\ \mathbf{b}_{02} \\ \vdots \\ \mathbf{b}_{0s} \end{bmatrix} \quad (15)$$

Each row corresponds to values of the input interval variables $\mathbf{b}_{0j}$ for which the deterministic model is going to be run, yielding the set of $s$ observed responses $\{w(\mathbf{b}_1), w(\mathbf{b}_2), \ldots, w(\mathbf{b}_s)\}$.

Thus, the initial training dataset is:

$$D_t = \{[\mathbf{b}_1, \mathbf{b}_2, \ldots, \mathbf{b}_s], [w(\mathbf{b}_1), w(\mathbf{b}_2), \ldots, w(\mathbf{b}_s)]\} \quad (16)$$

Therefore, the GP is built considering $D = D_t$. Alternative sampling methods [21,35], e.g. the Latin Hypercube sampling, could be also implemented to obtain the initial training dataset.

## 2.5 Selecting the elements of the Enhanced Training Dataset

Having defined the GP regression model of the objective function $w(\mathbf{b})$, the Bayesian optimization framework [31,32] can be readily employed for the selection of the next point $\hat{\mathbf{b}}_j$ at which the $w(\mathbf{b})$ has to be evaluated in order to improve the most the prediction of the maximum of the objective function. In particular, the next point $\hat{\mathbf{b}}_j$ is evaluated by formulating an optimization problem in the form [28,31,32]:

$$\hat{\mathbf{b}}_j = \arg\max_{\underline{\mathbf{b}} \leq \mathbf{b} \leq \overline{\mathbf{b}}} \left[ AF(\mathbf{b}) \right] \quad (17)$$

Where $AF(\mathbf{b})$ represent the so-called Acquisition Function (AF), that is an inexpensive function of the GP regression model which quantifies how desirable is to evaluate the objective function at $\mathbf{b}_j$ [28,31,32]. The AF is chosen to respect two criteria [28,31,32]: (i) Exploitation which suggests that $\hat{\mathbf{b}}_j$ should be located at regions with low variance values close to the current optimum value; (ii) Exploration which suggests that $\hat{\mathbf{b}}_j$ located at unexplored regions with high variance. Eq. (17) can be particularized for tackling the evaluation of the next point when predicting the LB and the UB of the response variable of interest, $\hat{\mathbf{b}}_{LB}$ and $\hat{\mathbf{b}}_{UB}$ respectively, so that

$$\hat{\mathbf{b}}_{UB} = \arg\max_{\underline{\mathbf{b}} \leq \mathbf{b} \leq \overline{\mathbf{b}}} \left[ AF^{(+)}(\mathbf{b}) \right] \quad (18)$$

$$\hat{\mathbf{b}}_{LB} = \arg\max_{\underline{\mathbf{b}} \leq \mathbf{b} \leq \overline{\mathbf{b}}} \left[ AF^{(-)}(\mathbf{b}) \right] \quad (19)$$

where the notation $AF^{(+)}(\mathbf{b})$ and $AF^{(-)}(\mathbf{b})$ is used to distinguish the formulation of the AF to be implemented in the two optimization problems in Eq. (1).

Many AFs have been proposed in the literature to balance the Exploitation and Exploration. An exhaustive review and comparison of the most known and used AFs can be found in references [32, 49-52]. Three AFs are explored in this paper: Probability of Improvement [21,33,37], Expected Improvement [28,31,32,34,37] and Confidence Bounds [38,39]. The AFs are often multimodal, and for complex problems, they might be non-convex and high-dimensional. Therefore, their maximization can be a challenging task [32]. Since the AF is relatively inexpensive to evaluate compared to the black-box response function, the resolution of the optimization problem in Eq. (18) and Eq. (19) is usually addressed with alternative optimization strategies, including gradient-based methods and variations of random search [31, 32]. In reference [29], closed-form expressions of the derivatives of the Expected Improvement AF with respect to the parameters $\mathbf{b}$ were used to tackle the AF optimization [29]. This approach might be prone to numerical errors related to calculation of the inverse of the training points covariance matrix and of the derivatives of the training and test points covariance matrix, especially in high-dimensions.

### 2.5.1 Probability of Improvement (PI)

The Probability of Improvement is based on the definition of the so-called Improvement function [21,33,37]. The improvement function with respect to the minimum value of the vector of observed responses $w_{min} = \min[\mathbf{w}_0]$ is defined as [21,33,37]:

$$I^{(-)}(\mathbf{b}) = \begin{cases} w_{min} - w(\mathbf{b}) & \text{if } w(\mathbf{b}) < w_{min} \\ 0 \end{cases} \quad (20)$$

So that values of the $w(\mathbf{b})$ higher than the current observed minimum yield zero improvement. Similarly, the improvement function with respect to $w_{max} = \max[\mathbf{w}_0]$ is written as [21,33,37]:

$$I^{(+)}(\mathbf{b}) = \begin{cases} w(\mathbf{b}) - w_{max} & \text{if } w(\mathbf{b}) > w_{max} \\ 0 \end{cases} \quad (21)$$

For each value of $\mathbf{b}_j$, the Improvement function is a linear combination of a deterministic value (the current optimum value) and of the GP regression model at $\mathbf{b}_j$ with predictive marginal mean $m(\mathbf{b})$ and variance $\sigma^2(\mathbf{b})$ (defined in Eq. (9) ad Eq. (10), respectively). Therefore, the probability density function of improvement is expressed as:

$$p\left(I^{(-)}(\mathbf{b})\right) \sim N\left(w_{min} - m(\mathbf{b}), \sigma^2(\mathbf{b})\right); \quad (22)$$

$$p\left(I^{(+)}(\mathbf{b})\right) \sim N\left(m(\mathbf{b}) - w_{max}, \sigma^2(\mathbf{b})\right) \quad (23)$$

Then the so-called Probability of Improvement (PI), for the minimization $PI^{(-)}(\mathbf{b})$ and maximization $PI^{(+)}(\mathbf{b})$ problems, is given by:

$$PI^{(\bullet)}(\mathbf{b}) = \int_0^\infty p\left(I^{(\bullet)}(\mathbf{b})\right) dI^{(\bullet)} = \Phi\left(\gamma^{(\bullet)}(\mathbf{b})\right) \quad (24)$$

With $\Phi(\bullet)$ being the cumulative distribution function of the standard normal and:

$$\gamma^{(+)}(\mathbf{b}) = \left(\frac{m(\mathbf{b}) - w_{\max}}{\sigma(\mathbf{b})}\right) \qquad \gamma^{(-)}(\mathbf{b}) = \left(\frac{w_{\min} - m(\mathbf{b})}{\sigma(\mathbf{b})}\right) \qquad (25)$$

The PI AF is more prone to select points favoring Exploitation [21,33,37]. Indeed, it favors sampling at locations with less uncertainty to yield an improvement over the current best observation [32,33]. This behavior can be observed in the results of the first numerical application shown in Figure 8(a).

### 2.5.2 Expected Improvement (EI)

The Expected Improvement (EI) was introduced by Močkus [34] and it is used as AF in the Efficient Global Optimization algorithm developed by Jones et al. [28]. The EI AF for the minimization $EI^{(-)}(\mathbf{b})$ and maximization $EI^{(+)}(\mathbf{b})$ problems can be defined as [28,34]:

$$EI^{(\bullet)}(\mathbf{b}) = \int_0^\infty I^{(\bullet)} p\left(I^{(\bullet)}(\mathbf{b})\right) dI^{(\bullet)}. \qquad (26)$$

The integrals in Eq. (26) can be solved by parts leading to the following closed-form expressions [21]:

$$EI^{(\bullet)}(\mathbf{b}) = \begin{cases} \sigma(\mathbf{b})\left\{\gamma^{(\bullet)}(\mathbf{b})\Phi\left[\gamma^{(\bullet)}(\mathbf{b})\right] + N\left[\gamma^{(\bullet)}(\mathbf{b});0,1\right]\right\} & \text{if } \sigma(\mathbf{b}) > 0; \\ 0 & \text{if } \sigma(\mathbf{b}) = 0, \end{cases} \qquad (27)$$

The EI AF is positive quantity that takes the value of zero at the observed responses. This is an important property which suggests that we cannot expect an improvement at points at which the model has been run [28,34]. In [28,34] it was shown that the EI yields a good balance between Exploitation and Exploration. This behavior can be observed in the results of the first numerical application shown in Figure 8(b). The performance of the EI AF has been investigated in several references [37, 52, 53]. Generalizations of this AF can be found in [54]. This AF has been used for assessing response bounds in engineering applications [26,27, 29].

### 2.5.3 Confidence Bounds (CBs)

The Confidence Bounds (CBs) are another type of AF [36,37]. The Lower Confidence Bound (LCB) and the Upper CB (UCB) are expressed as [38,39]:

$$LCB(\mathbf{b}, \chi) = -(m(\mathbf{b}) - \chi\sigma(\mathbf{b})), \tag{28}$$

$$UCB(\mathbf{b}, \chi) = m(\mathbf{b}) + \chi\sigma(\mathbf{b}). \tag{29}$$

with $\chi \geq 0$. The parameter $\chi$ balances the trade-off between Exploitation and Exploration. In particular, the lower is the value of $\chi$, the more Exploitation is obtained. According to [38], $\chi = 2$ yield a good compromise. More advanced strategies can be implemented to tune $\chi$, as described in [39]. Results obtained for the first numerical applications shown in Figure 8(c) with $\chi = 2$ indeed display a good balance between Exploration and Exploitation.

## 2.6 Enhanced training dataset

The new point $\hat{\mathbf{b}}_{LB}$ is selected by using the GP regression model obtained with the initial training dataset and by considering one AF for improving the LB estimate. Similarly, $\hat{\mathbf{b}}_{UB}$ is selected by using the same GP regresion model and considering on AF for improving the UB estimate. This means that each AF does not consider the improvement that would be achieved by selecting simultaneously or sequentially the two points, but it considers the effect of adding one observation only to the training dataset of the GP regression model. Once $\hat{\mathbf{b}}_{LB}$ and $\hat{\mathbf{b}}_{UB}$ are identified, the two additional deterministic analyses are carried out to yield $w(\hat{\mathbf{b}}_{LB}), w(\hat{\mathbf{b}}_{UB})$. These values are used to create the so-called enhanced training dataset.

In Approach A, two separate enhanced training datasets are generated at each iteration:

$$D_{UB} = \left(\left[\mathbf{b}_1, \ldots, \mathbf{b}_{0n_0}, \hat{\mathbf{b}}_{UB}\right], \left[w(\mathbf{b}_1), \ldots, w(\mathbf{b}_{0n_0}), w(\hat{\mathbf{b}}_{UB})\right]\right) \tag{30}$$

$$D_{LB} = \left(\left[\mathbf{b}_1, \ldots, \mathbf{b}_{0n_0}, \hat{\mathbf{b}}_{LB}\right], \left[w(\mathbf{b}_1), \ldots, w(\mathbf{b}_{0n_0}), w(\hat{\mathbf{b}}_{LB})\right]\right) \tag{31}$$

Then $D = D_{UB}$ is used in the next step of the iterative procedure for building a new GP that is then used exclusively for evaluating the UB. This means that only one element is added to $D_{UB}$ at each iteration. Similarly, $D = D_{LB}$ is used for building a separate GP that is used exclusively for determining the LB. Therefore, the two optimization problems are treated separately.

Whereas in Approach B a single enhanced training dataset is considered:

$$D_E = \left(\left[\mathbf{b}_1,\ldots,\mathbf{b}_{0n_0},\hat{\mathbf{b}}_{LB},\hat{\mathbf{b}}_{UB}\right],\left[w(\mathbf{b}_1),\ldots,w(\mathbf{b}_{0n_0}),w(\hat{\mathbf{b}}_{UB}),w(\hat{\mathbf{b}}_{LB})\right]\right) \quad (32)$$

So that for the next step of the iterative procedure $D = D_E$. In this approach, at each iteration a single GP is used for estimating the LB and UB. Therefore, $\hat{\mathbf{b}}_{LB}$ and $\hat{\mathbf{b}}_{UB}$ are evaluated by considering the same GP and two independent AFs, one for improving the LB and another for the UB estimates indepentently. That is, each AF will take into account the effect of adding one point only to the training dataset. However, two elements $w(\hat{\mathbf{b}}_{LB}), w(\hat{\mathbf{b}}_{UB})$ are going to be added to the enhanced training dataset at the end of each iteration.

## 2.7 Stopping Criteria

The stopping criteria define the conditions under which the iterations are stopped. The most obvious stopping criterion is the computational budget. This is because in engineering applications, the number of simulations that it is possible to run is usually limited by time and/or cost constraints. However, a pre-fixed computational budget may lead to stopping before the actual response bounds have been correctly identified. Alternatively, late stopping might occur, which would result in the unnecessary allocation of computational resources.

Within Bayesian optimization, it is common to use alternative stopping criteria [31,32]. One popular stopping criterion [27, 28, 29, 37, 38] is expressed in terms of the AF being smaller than a specified tolerance value $\Delta$ [37,38]:

$$\left|\max_{\underline{\mathbf{b}}\leq\mathbf{b}\leq\overline{\mathbf{b}}}\left[AF(\mathbf{b})\right]\right| < \Delta. \quad (33)$$

For the EI AF, it is suggested to set $\Delta$ to $10^{-2}$ [37]. For the CBs, $\Delta$ would correspond either the minimum or the maximum of the observed responses ($w_{\min} = \min[\mathbf{w}_0]$ or $w_{\max} = \max[\mathbf{w}_0]$, respectively) [38]. Both criteria are going to be considered and verified at the end of each iteration step. It is worth noting that the satisfaction of the stopping criterion in Eq. (33) does not necessarily guarantee a certain accuracy on the response bounds estimate, as discussed in what follows.

## 2.8 Response bound estimates

Once the stopping criteria are met, the latest GP regression model can be used to obtain the prediction on the response LB and UB in terms of:

(i) the minimum and maximum of the training points $(w_{\min}, w_{\max})$ obtained at

$$\mathbf{b}_{w_{\min}} = \arg\min_{\mathbf{B}_0}\left(w(\mathbf{b}_{0j})\right) \qquad \mathbf{b}_{w_{\max}} = \arg\max_{\mathbf{B}_0}\left(w(\mathbf{b}_{0j})\right) \tag{34}$$

(ii) the minimum and/or maximum of the mean function of the posterior distribution, $m(\mathbf{b}_{LB}), m(\mathbf{b}_{UB})$ obtained at

$$\mathbf{b}_{LB} \approx \arg\min_{\mathbf{b}}\left(m(\mathbf{b})\right) \qquad \mathbf{b}_{UB} \approx \arg\max_{\mathbf{b}}\left(m(\mathbf{b})\right) \tag{35}$$

These are standard ways of expressing the results obtained with Bayesian Optimization [31,32], and (i) is usually preferred in engineering applications [26, 27, 29]. Nonetheless, the mean $m(\mathbf{b})$ can be interpreted as a point estimate of $w(\mathbf{b})$ for a fixed $\mathbf{b}$, and using the frequentist interpretation, the credible interval $m(\mathbf{b}) \pm 2\sigma(\mathbf{b})$ can be interpreted as the 95% confidence interval – that is the interval in which there is a 95% probability to find the true function $w(\mathbf{b})$. The confidence bounds returned at the minimum and/or maximum of the mean function can be used to quantify the remaining epistemic uncertainty caused by not having evaluated the simulation at that point. Therefore, each bound obtained with Approach A or Approach B can be expressed as an interval:

$$\underline{w} = \left[m(\mathbf{b}_{LB}) - 2\sigma(\mathbf{b}_{LB}), m(\mathbf{b}_{LB}) + 2\sigma(\mathbf{b}_{LB})\right] \qquad \overline{w} = \left[m(\mathbf{b}_{UB}) - 2\sigma(\mathbf{b}_{UB}), m(\mathbf{b}_{UB}) + 2\sigma(\mathbf{b}_{UB})\right]$$

$$\tag{36}$$

The lower bound of the LB response, $\underline{w}^{LB}$, and the upper bound of the UB response, $\overline{w}^{UB}$, are, respectively:
$$\underline{w}^{LB} = m(\mathbf{b}_{LB}) - 2\sigma(\mathbf{b}_{LB}) \qquad \overline{w}^{UB} = m(\mathbf{b}_{UB}) + 2\sigma(\mathbf{b}_{UB}) \tag{37}$$

These bounds are more conservative than those yielded by considering the deterministic observations, and not necessarily with respect to the true response bounds. In particular, because of the way the GP regression model has been built, $\overline{w}^{UB} \geq w_{\max}$ and $\underline{w}^{LB} \leq w_{\min}$. Therefore the $\overline{w}^{UB}$ is over-predicting the UB obtained with the deterministic observations ($w_{\max}$) when $\mathbf{b}_{UB} \neq \mathbf{b}_{w_{\max}}$, and otherwise $w_{\max} = m(\mathbf{b}_{UB})$. Similarly, the $\underline{w}^{LB}$ is under-predicting the LB obtained with the deterministic observations ($w_{\min}$) when $\mathbf{b}_{LB} \neq \mathbf{b}_{w_{\min}}$, and otherwise $w_{\min} = m(\mathbf{b}_{LB})$.

Approach A yields two separate posterior mean functions, while only one posterior mean function is evaluated in Approach B. Consequently, for a fixed number of iterations, the predicted bounds obtained with the two approaches would be different. This is because within Approach B, the simulations included in the enhanced training dataset are not selected to improve only one of the response bound estimates. Therefore, the enhanced training dataset built iteratively within Approach B would also enhance the overall regression model by exploring regions of the objective function that would not be explored otherwise. This would reduce the possibility of the optimiser being stuck in a local optimum point and, depending on the problem investigated, might lead to a reduction of the number of simulations needed to yield sufficiently accurate bound estimates according to the chosen stopping criterion.

2.9 **Metrics for verifying if the bound estimates are satisfactory**

It is particularly important to verify that the bound estimates obtained after a stopping criterion has been reached are satisfactory or otherwise. This information is not entirely carried out by the bounds in Eq. (36) and Eq. (37). Returning a point estimate for one bound merely means that a simulation has been evaluated at that point, not that the true maximum or minimum of the response has been estimated. Since the underlying response at every point of the objective function is not known, it is important to define metrics which can be used to verify if the bound estimates can be considered satisfactory or if more simulations should be evaluated. Two metrics are proposed.

Metric 1: Let us consider the last point selected by the acquisition function (without running an additional simulation) $\hat{\mathbf{b}}_j\big|_{LI}$, where the subscript $LI$ indicates the value obtained at the Last Iteration, and the value $\mathbf{b}_{LB}$ (or $\mathbf{b}_{UB}$) at which the current LB is predicted (as obtained by using Eq. (35)). In the interval variables space, $\hat{\mathbf{b}}_j\big|_{LI,LB}$ and $\mathbf{b}_{LB}$, and similarly $\hat{\mathbf{b}}_j\big|_{LI,UB}$ and $\mathbf{b}_{UB}$, can considered as position vectors. The Euclidian distance of these two vectors can be then considered:

$$z_{LB,i} = \left\|\left(\mathbf{b}_{LB} - \hat{\mathbf{b}}_j\big|_{LI,LB}\right)\cdot \mathbf{e}_i\right\|_2 ; \tag{38}$$

$$z_{UB,i} = \left\|\left(\mathbf{b}_{UB} - \hat{\mathbf{b}}_j\big|_{LI,UB}\right)\cdot \mathbf{e}_i\right\|_2 \tag{39}$$

Where $\mathbf{e}_i$ is the unit vector along the length of each interval variable so that $i=1,2,\ldots,r$ and the dot indicates the scalar product. If the distance in Eq. (38) or Eq. (39) is large, the bound estimate obtained might be inaccurate, since the true response bound might occur in a part of the input interval variables domain that has not been sufficiently explored. As a rule-of-thumb, a distance greater than 2% of the length of the corresponding input interval variable should be investigated. Therefore, it should be verified that:

$$z_{LB,i} < 0.02\left|b_i^{\text{int}}\right| \tag{40}$$

$$z_{UB,i} < 0.02\left|b_i^{\text{int}}\right| \tag{41}$$

Moreover, the confidence interval obtained with the GP at the point identified by the AF, that is $m\left(\hat{\mathbf{b}}_j\big|_{LI}\right) \pm 2\sigma\left(\hat{\mathbf{b}}_j\big|_{LI}\right)$ should be such that:

$$m\left(\hat{\mathbf{b}}_j\big|_{LI,LB}\right) - 2\sigma\left(\hat{\mathbf{b}}_j\big|_{LI,LB}\right) > \underline{w}^{LB} \tag{42}$$

$$m\left(\hat{\mathbf{b}}_j\big|_{LI,UB}\right) + 2\sigma\left(\hat{\mathbf{b}}_j\big|_{LI,UB}\right) < \overline{w}^{UB} \tag{43}$$

where $\underline{w}^{LB}$ and $\overline{w}^{UB}$ are defined in Eq. (37). A warning flag should be raised if Eq. (40) and Eq. (42) and/or Eq. (41) and Eq. (43) are not simultaneously met. It is important to note that these conditions might not be satisfied for Approach A, even when the correct bound has been identified. After finding a local optimum, the AF might suggest to explore regions of the objective function which are yet unexplored to identify either the true maximum or the minimum of the objective function. For example, it is possible to observe this effect in the results obtained in the numerical application for iteration 5 of Figure 8 (b) and iteration 10 of Figure 8(c). Approach B would be less susceptible to this, given that a single GP is used for identifying both the maximum and the minimum of the response variable, and therefore, more regions of the response function would be explored.

Metric 2: Let us consider the Euclidian distance between $\mathbf{b}_{w_{min}}$ (at which the minimum value of the vector of observed responses $w_{min}$ is obtained) and $\mathbf{b}_{LB}$ along the $\mathbf{e}_i$, and in a similar fashion between the components of $\mathbf{b}_{w_{max}}$ and $\mathbf{b}_{UB}$, so that:

$$\left\|\left(\mathbf{b}_{LB} - \mathbf{b}_{w_{min}}\right) \cdot \mathbf{e}_i\right\|_2 < 0.02 \left|b_i^{int}\right| \tag{44}$$

$$\left\|\left(\mathbf{b}_{UB} - \mathbf{b}_{w_{max}}\right) \cdot \mathbf{e}_i\right\|_2 < 0.02 \left|b_i^{int}\right| \tag{45}$$

These conditions are important to ensure that the bound estimates are obtained for values of the interval variable near where $w_{min}, w_{max}$ have been observed. However, it is also important to compare the magnitude of the mean function of the posterior distribution obtained at these points. As a rule-of-thumb, a variation in magnitude of the mean function of the posterior distribution higher than 5% might indicate a highly non-linear response function. Therefore, it should be simultaneously verified that:

$$\left|m(\mathbf{b}_{LB}) - m(\mathbf{b}_{w_{min}})\right| < 0.05 \times \left|m(\mathbf{b}_{LB})\right| \tag{46}$$

$$\left|m(\mathbf{b}_{UB}) - m(\mathbf{b}_{w_{max}})\right| < 0.05 \times \left|m(\mathbf{b}_{UB})\right| \tag{47}$$

It is worth noting that the conditions expressed in Eq. (46) and Eq. (47) could be met even when Eq. (44) and Eq. (45) do not hold. This is a clear warning that although the magnitude of the bounds currently predicted with the GP is close to a deterministic point, the response bounds might be located in different regions of the objective function. It is important to stress here, that satisfying only these conditions (Eq. (44) –Eq. (47)) does not imply that the true bounds have been obtained.

In conclusions, if any of the conditions expressed in Eq. (40) – Eq. (47) are not met, further simulations should be carried out to reduce potentially large inaccuracies in each bound estimate.

### 3. Summary of steps for Approach A
1. Build the initial training dataset:
    a. Select the initial $s$ combinations of interval variables $[\mathbf{b}_1, \mathbf{b}_2, \ldots, \mathbf{b}_s]$
    b. Evaluate the response variable at those points $[w(\mathbf{b}_1), w(\mathbf{b}_2), \ldots, w(\mathbf{b}_s)]$ by running $s$ simulations

c. Construct the training dataset: $D_t = \{[\mathbf{b}_1, \mathbf{b}_2, \ldots, \mathbf{b}_s], [w(\mathbf{b}_1), w(\mathbf{b}_2), \ldots, w(\mathbf{b}_s)]\}$
   d. Set $D = D_t$
2. Build the GP regression model:
   a. Use the available training dataset $D$
   b. Set the size of the test dataset $n_*$ and select the $\mathbf{B}_* = [\mathbf{b}_{*1}, \ldots, \mathbf{b}_{0n_*}]$
   c. Set the mean functions $\mu_* = 0; \mu_0 = 0$
   d. Select the kernel function: exponential of a weighted distance function
   e. Evaluate the parameters of the kernel function by maximizing the log-marginal likelihood of the observed data
   f. Evaluate the mean and covariance functions of the marginal posterior predictive distribution by using Eq. (9) and Eq. (10)
3. For UB Select the new simulation and build the enhanced the training dataset by
   a. Selecting the AF to be used: PI, EI and CBs (see section 2.3)
   b. Solving the optimization problems for $AF^{(+)}(\mathbf{b})$ (Eq. (18)) to yield $\hat{\mathbf{b}}_{UB}$
   c. Performing the simulations for $\hat{\mathbf{b}}_{UB}$ to yield $w(\hat{\mathbf{b}}_{UB})$
   d. Augmenting the training dataset for the UB

   $$D_{UB} = \left( \left[ \mathbf{b}_1, \ldots, \mathbf{b}_{0n_0}, \hat{\mathbf{b}}_{UB} \right], \left[ w(\mathbf{b}_1), \ldots, w(\mathbf{b}_{0n_0}), w(\hat{\mathbf{b}}_{UB}) \right] \right)$$

   e. Set $D = D_{UB}$
4. For UB: Repeat steps 2-3 until the stop criteria for UB are met:
   a. Check the computational budget
   b. Check the stopping criterion on AF (Eq. (33))
5. For UB return
   a. $\hat{\mathbf{b}}_{UB}$ (Eq. (35)) and $\overline{w} = m(\hat{\mathbf{b}}_{UB}) \pm 2\sigma(\hat{\mathbf{b}}_{UB})$ (Eq. (36))
   b. $\mathbf{b}_{w_{max}}$ and $m(\mathbf{b}_{w_{max}})$ (Eq. (34))
   c. $\hat{\mathbf{b}}_j\big|_{LI}$ and $m(\hat{\mathbf{b}}_j\big|_{LI}) \pm 2\sigma(\hat{\mathbf{b}}_j\big|_{LI})$
   d. Verify if the conditions in Eq. (41), Eq. (43), Eq. (45), Eq. (47) are met

In parallel:

3b. For LB: Build the enhanced the training dataset by

   a. Selecting the AF to be used: PI, EI and CBs (see section 2.3)
   b. Solving the optimization problems for $AF^{(-)}(\mathbf{b})$ (Eq. (19)) to yield $\hat{\mathbf{b}}_{LB}$
   c. Performing the simulations for $\hat{\mathbf{b}}_{LB}$ to yield $w(\hat{\mathbf{b}}_{LB})$
   d. Augmenting the training dataset for the LB

   $$D_{LB} = \left( \left[ \mathbf{b}_1, \ldots, \mathbf{b}_{0n_0}, \hat{\mathbf{b}}_{LB} \right], \left[ w(\mathbf{b}_1), \ldots, w(\mathbf{b}_{0n_0}), w(\hat{\mathbf{b}}_{LB}) \right] \right)$$

e. Set $D = D_{LB}$

4b. For LB: Repeat steps 2-3b until the stop criteria for LB are met:

   f. Check the computational budget
   g. Check the stopping criterion on AF (Eq. (33))

5b. For UB return

   e. $\mathbf{b}_{LB}$ (Eq. (35)) and $\underline{w} = m(\mathbf{b}_{LB}) \pm 2\sigma(\mathbf{b}_{LB})$ (Eq. (36))
   f. $\mathbf{b}_{w_{\min}}$ and $m(\mathbf{b}_{w_{\min}})$ (Eq. (34))
   g. $\hat{\mathbf{b}}_j\big|_{LI}$ and $m(\hat{\mathbf{b}}_j\big|_{LI}) \pm 2\sigma(\hat{\mathbf{b}}_j\big|_{LI})$
   h. Verify if the conditions in Eq. (40), Eq. (42), Eq. (44), Eq. (46) are met

## 4. Summary of steps for Approach B

1. Build the initial training dataset:
   a. Select the *s* combinations of the interval variables $[\mathbf{b}_1, \mathbf{b}_2, \ldots, \mathbf{b}_s]$.
   b. Evaluate the response variable at those points $[w(\mathbf{b}_1), w(\mathbf{b}_2), \ldots, w(\mathbf{b}_s)]$ by running *s* simulations.
   c. Construct the training dataset: $D_t = \{[\mathbf{b}_1, \mathbf{b}_2, \ldots, \mathbf{b}_s], [w(\mathbf{b}_1), w(\mathbf{b}_2), \ldots, w(\mathbf{b}_s)]\}$
   a. Set $D = D_t$
2. Build the GP regression model:
   a. Use the available training dataset $D$
   b. Set the size of the test dataset $n_*$ and select the $\mathbf{B}_* = [\mathbf{b}_{*1}, \ldots, \mathbf{b}_{0n_*}]$
   c. Set the mean functions $\boldsymbol{\mu}_* = \mathbf{0}; \boldsymbol{\mu}_0 = \mathbf{0}$
   d. Select the kernel function: exponential of a weighted distance function
   e. Evaluate the parameters of the kernel function by maximizing the log-marginal likelihood of the observed data
   b. Evaluate the mean and covariance functions of the marginal posterior predictive distribution by using Eq. (9) and Eq. (10)
3. Select the new simulation and build the enhanced the training dataset by
   a. Selecting the AF to be used: PI, EI and CBs (see section 2.3)
   b. Solving independently two optimization problems using the same GP:
      i. for $AF^{(+)}(\mathbf{b})$ (Eq. (18)) to yield $\hat{\mathbf{b}}_{UB}$
      ii. for $AF^{(-)}(\mathbf{b})$ (Eq. (19)) to yield $\hat{\mathbf{b}}_{LB}$
   c. Performing the simulations for:
      i. $\hat{\mathbf{b}}_{UB}$ to yield $w(\hat{\mathbf{b}}_{UB})$

ii. $\hat{\mathbf{b}}_{LB}$ to yield $w(\hat{\mathbf{b}}_{LB})$

d. Augmenting the training dataset

$$D_E = \left(\left[\mathbf{b}_1,\ldots,\mathbf{b}_{0n_0},\hat{\mathbf{b}}_{LB},\hat{\mathbf{b}}_{UB}\right],\left[w(\mathbf{b}_1),\ldots,w(\mathbf{b}_{0n_0}),w(\hat{\mathbf{b}}_{LB}),w(\hat{\mathbf{b}}_{UB})\right]\right)$$

e. Set $D = D_E$

4. Repeat steps 2-3 until the stop criteria are met on both bounds:
    a. Check the computational budget
    b. Check the stopping criterion on AF (Eq. (33))
    c. If the stopping criteria are reached for one bound, the optimization for that bound can be stopped
5. Return:
    a. $\mathbf{b}_{UB}$ (Eq. (35)) and $\underline{w} = m(\mathbf{b}_{UB}) \pm 2\sigma(\mathbf{b}_{UB})$ (Eq. (36))
    b. $\mathbf{b}_{w_{max}}$ and $m(\mathbf{b}_{w_{max}})$ (Eq. (34))
    c. $\hat{\mathbf{b}}_j\big|_{LI}$ and $m(\hat{\mathbf{b}}_j\big|_{LI}) \pm 2\sigma(\hat{\mathbf{b}}_j\big|_{LI})$ for UB
    d. Verify if the conditions in Eq. (41), Eq. (43), Eq. (45) and Eq. (47) are met
    e. $\mathbf{b}_{LB}$ (Eq. (35)) and $\underline{w} = m(\mathbf{b}_{LB}) \pm 2\sigma(\mathbf{b}_{LB})$ (Eq. (36))
    f. $\mathbf{b}_{w_{min}}$ and $m(\mathbf{b}_{w_{min}})$ (Eq. (34))
    g. $\hat{\mathbf{b}}_j\big|_{LI}$ and $m(\hat{\mathbf{b}}_j\big|_{LI}) \pm 2\sigma(\hat{\mathbf{b}}_j\big|_{LI})$ for LB
    h. Verify if the conditions in Eq. (40), Eq. (42), Eq. (44), Eq. (46) are met

## 5. Numerical Applications

The proposed approaches are investigated and validated by considering two case studies. The first numerical application concerns the evaluation of the bounds of the maximum absolute value of the acceleration of a Single Degree of Freedom system with interval stiffness (one uncertain parameter) subject to a sine wave. This case study is used to investigate the effects of (i) non-monotonicity; (ii) choice of AFs; (iii) stopping criteria; (iv) size of initial training dataset. The assessment of the applicability of the proposed approaches to an expensive-to-evaluate model with many uncertainties, could lead to issues about reproducibility of the results by other users, as well as problems of drawing useful conclusions from the results. The deterministic model developed in reference [55] is chosen given that it provides a good trade-off between complexity and computational cost, and that it comes with a step-by-step guide for the model development with a FE commercial software. This second numerical application focuses on the evaluation of the bounds of the Sound Pressure Level obtained at a specific frequency and at one point inside the cavity of a plate-acoustic-cavity system driven by a unit harmonic force applied to the plate. Four uncertain parameters are introduced in the model developed in [55]. This numerical example is used to investigate further the influence of: (i) the size of the initial training dataset; (ii) the selected acquisition functions; (iii) the stopping criteria.

When applying Approach A and Approach B, the optimization problems were addressed with robust strategies which are less efficient than advanced optimization solvers. In particular, the Matlab function "patternsearch" [56] was used for the loglikelihood minimization (by using Eq. (13)) in order to determine the parameters of the kernel function, and the following ranges were specified: $1 \leq p_h \leq 2$, $0 \leq \theta_h \leq 2$. The optimization of the AFs was performed by using a brute-force search. The accuracy of Approach A and Approach B is assessed by direct comparison with the results provided by the Sub-Interval Method (SM) and Vertex Method (VM). It is worth noting that variances of the order of $10^{\wedge}(-10)$ were approximated to zero.

### 5.1 Single Degree of Freedom with a single interval uncertainty

Let us consider a mass-spring-damper system characterized by a mass $M = 1000 \,\text{kg}$, a viscous damper with $c = 1.98 \,\text{kNs/m}$ and an interval stiffness $k^{\text{int}} = [1715, 3185] \,\text{kN/m}$. The mass is subject to a time-varying input $f(t) = F_0 \sin(\omega_f t)$ with amplitude $F_0 = 100 \,\text{kN}$ and driving frequency $\omega_f = 4\pi \,\text{rads}^{-1}$ in the time-interval $0 \leq t \leq 0.5$ seconds, and it is subject to zero initial conditions. The dynamic response is governed by the well-known second-order differential equation [57]. The response variable of interest is the maximum of the absolute acceleration time history $|a(t,k)|$ obtained in the time window $0 \leq t \leq T_f$ with $T_f = 5\,\text{s}$ for a given realisation of $k$. The objective function is expressed as:

$$w(k) = \max_{0 \leq t \leq T_f} \left[ |a(t,k)| \right] \tag{46}$$

Two bounded constrained optimization problems (in the form of Eq. (1)) have to be solved to yield the Lower Bound (LB) $\underline{w}(k)$ and Upper Bound (UB) $\overline{w}(k)$ of the variable of interest. Each realization of $w(k)$ is obtained by solving numerically the governing differential equation for a fixed *k*. The Transition Matrix Method integration scheme [58] was used, and a time-step of $\Delta t = 10^{-3}$ s was considered.

The SM was applied by subdividing $k^{int}$ into *n*=300 uniform subintervals. The target response function $w(k)$, shown in Figure 3, was evaluated by performing 301 simulations. From Figure 3, it is possible to observe that SM yields very accurate predictions of the response UB and LB. The response bounds yielded by the VM (obtained by setting the *k* to its maximum and minimum value) are instead underestimating the UB and overestimating the LB. This is because $w(k)$ is not a monotonic function with respect to the interval variable *k*, as assumed in VM.

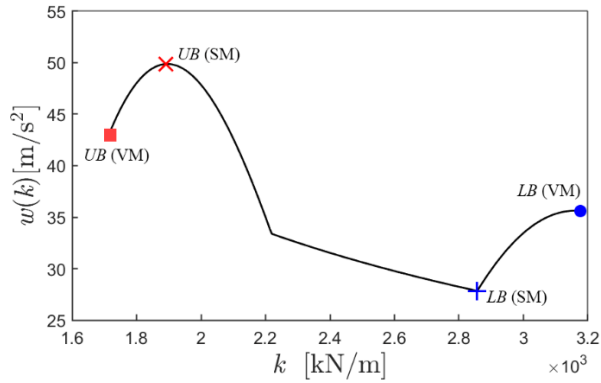

Figure 3. Response variable as a function of the interval variable. The markers "x" and "+" indicate the maximum and minimum of the response obtained with the Sub-Interval Method (SM), respectively. The markers "square" and "circle" indicate the maximum and minimum of the response obtained with the Vertex Method (VM), respectively.

In the next subsection, the results yielded by the SM and the VM approaches are compared when considering six different deviations of the uncertain variable with respect to its nominal value to investigate the non-monotonicity effects on the response bounds.

### 5.1.1 Non-monotonicity effects on the response bounds

Let us indicate $k^{int} = [\underline{k}, \overline{k}]$ where $\underline{k} = k_0(1 - \Delta b)$ and $\overline{k} = k_0(1 + \Delta b)$ with $k_0$ being the nominal value of the stiffness $k_0 = 2450 \, \text{kN/m}$ and $\Delta b$ describing the dimensionless deviation of the uncertain variable with respect to the nominal value. In Figure 4, the lower and upper bounds yielded by the VM and SM approaches for $\Delta b = 0.05:0.05:0.3$ are compared.

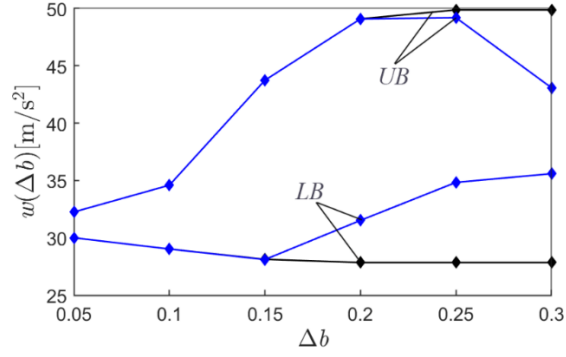

Figure 4. Response bounds obtained for six values of the dimensionless variation when using the Subinterval Method (black curves) and the Vertex Method (blue curves).

It is possible to observe that for each $\Delta b$, different UB and LB of the response are obtained. For $\Delta b = 0.15$, the interval variable $k$ is $[2082.5, 2817.5]$ kN/m. By inspection of Figure 3, it is possible to observe that in this range, and for values of $\Delta b \leq 0.15$, the response function $w(k)$ is a monotonic function with respect to $k$. Therefore the results obtained with the SM and the VM coincide. This is not the case for $\Delta b \geq 0.2$. The response bounds predicted by VM become inaccurate because of the non-monotonicity of $w(k)$ with respect to $k$. In particular, for $\Delta b = 0.3$ the VM overestimates the LB by 27.78% and underestimates the UB by 13.60%.

### 5.1.2 Initial training dataset

The interval is partitioned in two parts $\mathbf{B} = \left[\underline{k}, k_0, \overline{k}\right]^T$. Consequently, the training dataset is $D = \left(\left[\underline{k}, k_0, \overline{k}\right], \left[w(\underline{k}), w(k_0), w(\overline{k})\right]\right)$. In Figure 5, the posterior mean function (red curve) and the 95% confidence intervals $(2\sigma(k))$ obtained with a GP considering 3 training points and 298 test points is shown. These results are compared to the response function obtained with the SM (black curve), showing large differences in the estimates of the response bounds.

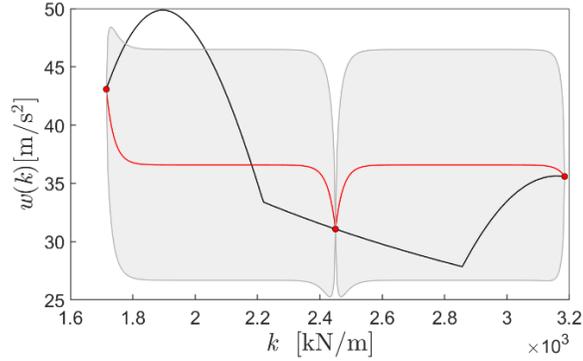

Figure 5. The mean function (red curve) and 95% confidence interval (shaded gray area) of the GP regression model of the response variable when using the initial training dataset (circle markers) compared to the target response function (black curve).

### 5.1.3 Approach A with fixed computational budget: investigation on AFs choice

The computational budget is fixed to 33 simulations, 3 of which were used for building the initial training dataset. The rest of the budget is equally shared for the estimation of the LB and of the UB. The results obtained with Approach A for different AFs are shown in Figure 6, 7 and 8. Figure 6 shows the maximum value of the AF and percentage error between the maximum/minimum of the posterior mean function predicted at each iteration with respect to the exact bounds obtained with SM. Figure 7 and Figure 8 show the GP mean function and 95% confidence interval obtained at several iterations for predicting the LB and UB, respectively. The blue and red vertical solid lines represent the location of the next candidate point identified by the chosen AF for minimizing (blue) and maximizing (red) the objective function.

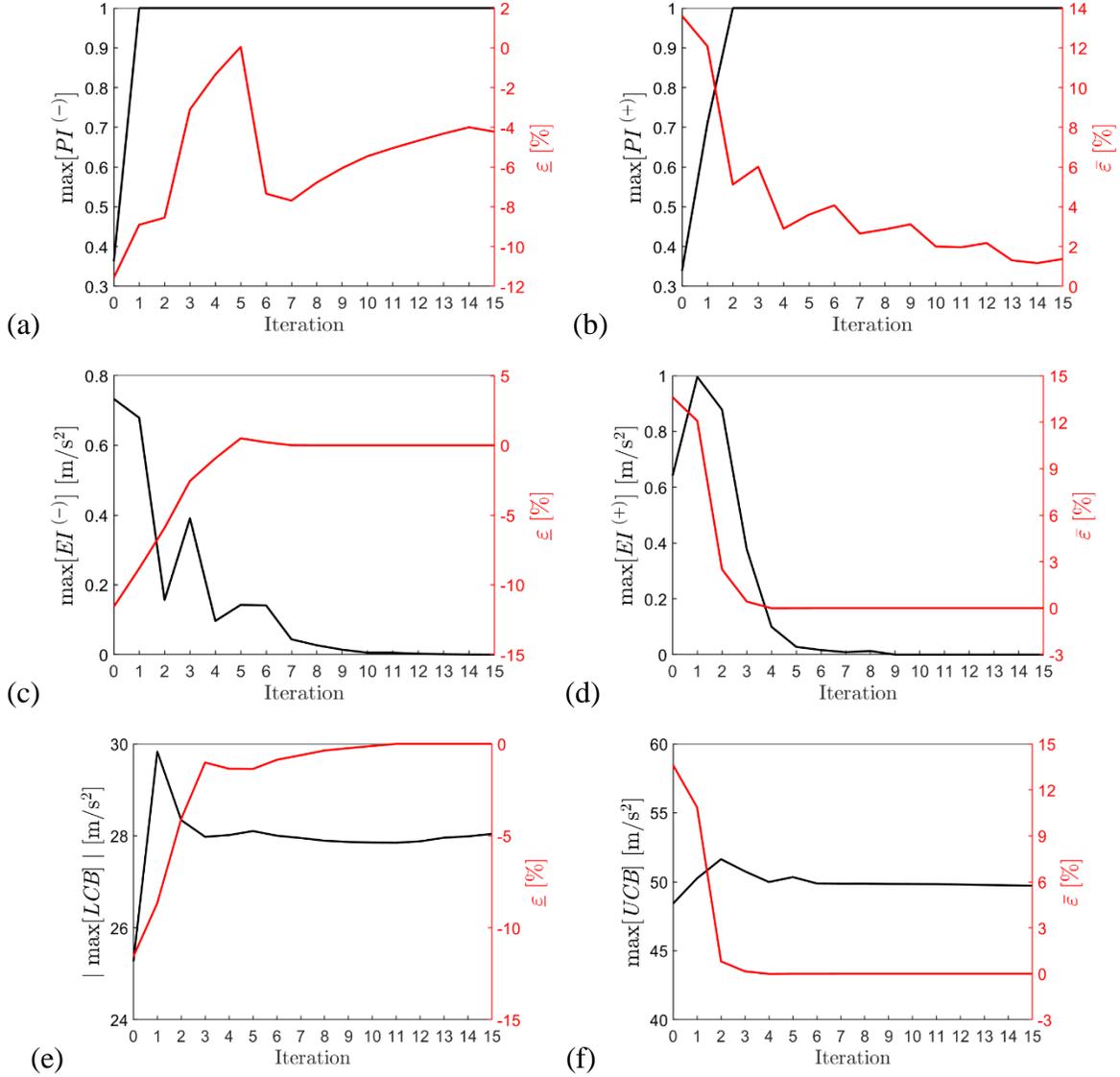

Figure 6. Maximum value of Probability of Improvement ((a) and (b)), Expected Improvement ((c) and (d)), Confidence Bounds ((e) and (f)), acquisition functions and percentage error in the bounds assessment for a given acquisition function as a function of the iteration number.

**PI AF.** From Figure 6a and 6b, it is possible to observe that as the number of iterations increases the maximum value of the PI does not decrease and its value is approximately one. Using the GP at the last iteration, an error $\underline{\varepsilon} = -4.22\%$ is observed on the LB, while the UB is estimated with $\bar{\varepsilon} = 1.38\%$. The fact that the PI AF favors Exploitation rather than Exploration is clearly observed in Figure 7a (LB) and Figure 8a (UB) which show the GP for the given enhanced training dataset, together with the location of the next point to be selected according to the AF (blue and red vertical lines for LB and UB, respectively). It is possible to observe that at each iteration the AF selects points very close to the optima of the deterministic observations.

**EI AF.** In Figure 6c and 6d, it is possible to notice that as the number of iterations increases the maximum value of the EI does not decrease monotonically. At the last iteration, $\underline{\varepsilon} = 0\%$ and $\overline{\varepsilon} = 1.706 \times 10^{-3}\%$. Therefore, in this case of study, EI was yielding a better approximation of the bounds compared to the PI AF. From Figure 7b and Figure 8b it is possible to observe that the value of the interval variable selected by the maximization of EI are progressively located around the exact locations of the LB and UB, but also explore other response regions once the maxima has been identified (as shown in iterations 5, 10 and 15 of Figure 8b). As anticipated, the EI AF yields better trade-off between the Exploitation and Exploration compared to the PI AF.

**CBs AF.** From Figure 6e and 6f it is possible to observe that as the number of iterations increases the maximum value of the CBs tends to be constant and the percentage errors at the last iteration are $\underline{\varepsilon} = 0\%$ and $\overline{\varepsilon} = 0\%$, that is, the bounds are exactly identified. From Figure 7c and Figure 8c, it is possible to notice that the values selected by the maximization of the CBs are progressively located around the exact LB (see Figure 7c) and UB (Figure 8c), but also explore other response regions once the maximum has been identified (as shown in iterations 5, 10 and 15 of Figure 8c).

In summary, the use of the EI and CBs as AFs showed an excellent performance in terms of the choice of the location of sampling points and the final accuracy for a fixed number of iterations. While the PI AF displayed a poor performance. Considering these results, hereinafter further investigations will be carried out comparing only the EI and CBs.

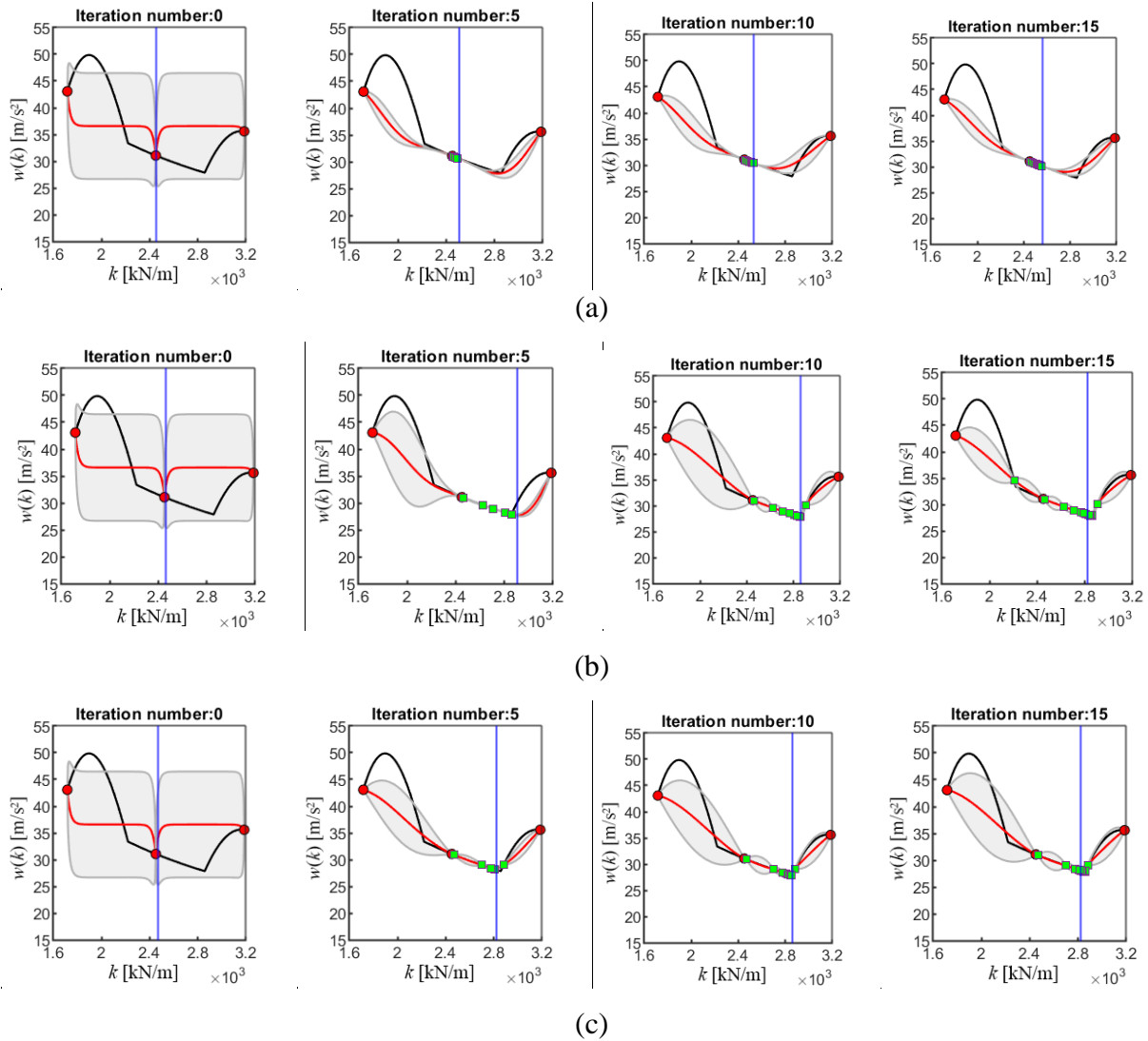

Figure 7. Evolution of the mean function (red curve) and 95% confidence interval (shaded gray area) of the GP regression model for evaluating the LB of the response by including the selected simulation results (circle and square markers) at iteration number 0,5,10 and 15 when using the acquisition function PI (a), EI (b) and CBs (c). The results are compared to the SM results (black curve). The blue vertical solid lines represent the location of the next candidate point for improving the LB prediction using the selected AF.

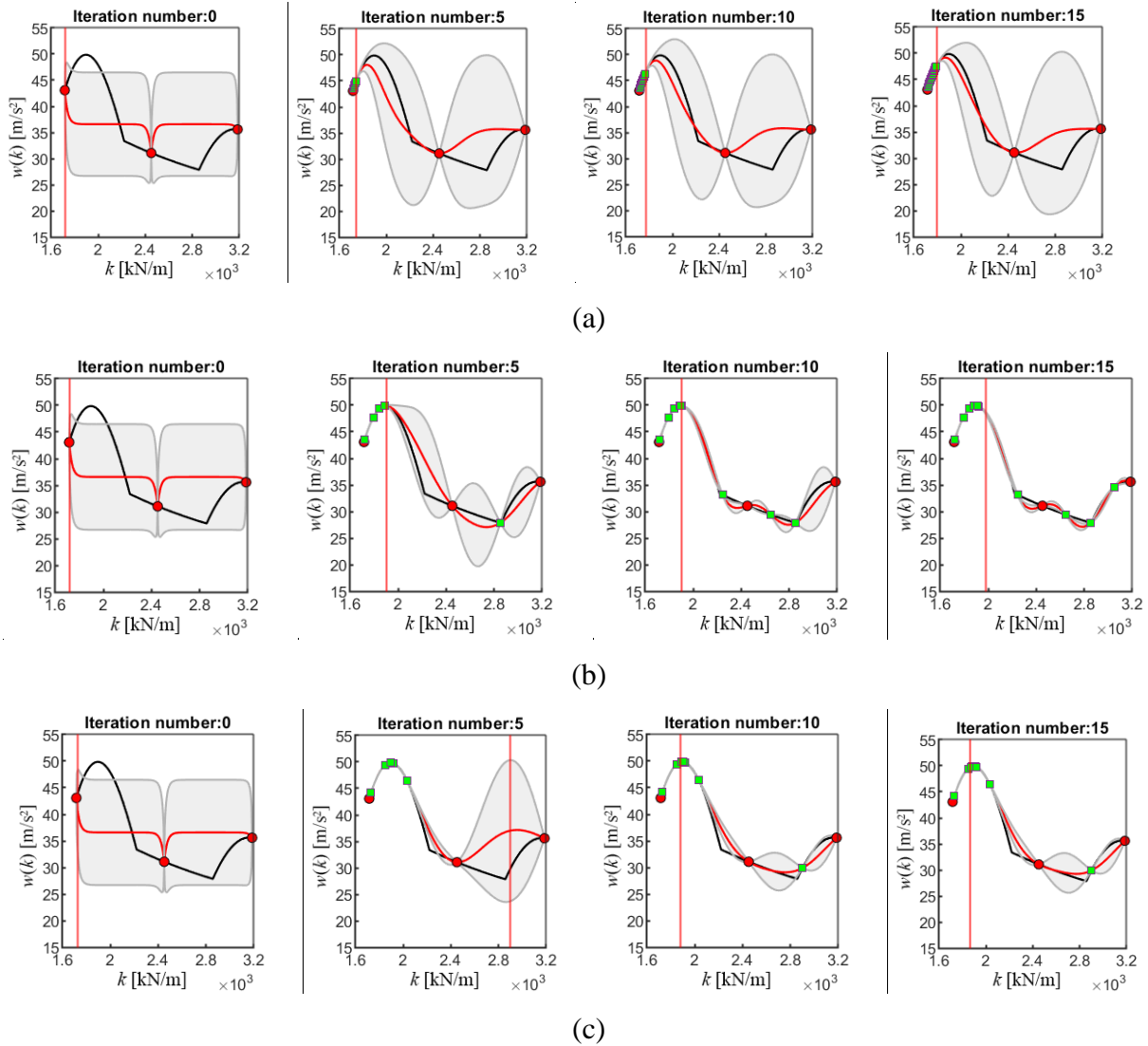

Figure 8. Evolution of the mean function (red curve) and confidence interval (shaded gray area) of the GP regression model for evaluating the UB of the response by including the selected simulation results (circle and square markers) at iteration number 0, 5, 10 and 15 when using the acquisition function PI (a), EI (b) and CBs (c). The results are compared to the SM results (black curve). The red vertical solid lines represent the location of the next candidate point for improving the UB prediction using the selected AF.

### 5.1.4 Approach A with stopping criterion on AF: investigation on AFs choice

**EI AF.** For the EI AF stopping criterion, $\Delta = 10^{-2}$ m/s$^2$ is considered. Figures 9a and 9b show respectively $\max\left[EI^{(-)}\right]$ and $\max\left[EI^{(+)}\right]$ at each iteration along with the percentage error $\underline{\varepsilon}[\%]$ and $\bar{\varepsilon}[\%]$. The stopping criterion for the $\max\left[EI^{(-)}\right]$ is achieved at the 10th iteration with an error on the LB estimate of $\underline{\varepsilon} = 0\%$. The stopping condition for $\max\left[EI^{(+)}\right]$ is obtained at the 7th iteration with an error on the UB estimate of $\bar{\varepsilon} = 1.706 \times 10^{-3}\%$. The total number of simulations to be evaluated is 20 (3 initials, 10 for LB and 7 for UB), which is lower than the total computational budget available. It is important to note that Figure 9a coincides with Figure 6c (up to the 10$^{th}$ iteration) and Figure 9b coincides with Figure 6d (up to the 7$^{th}$ iteration).

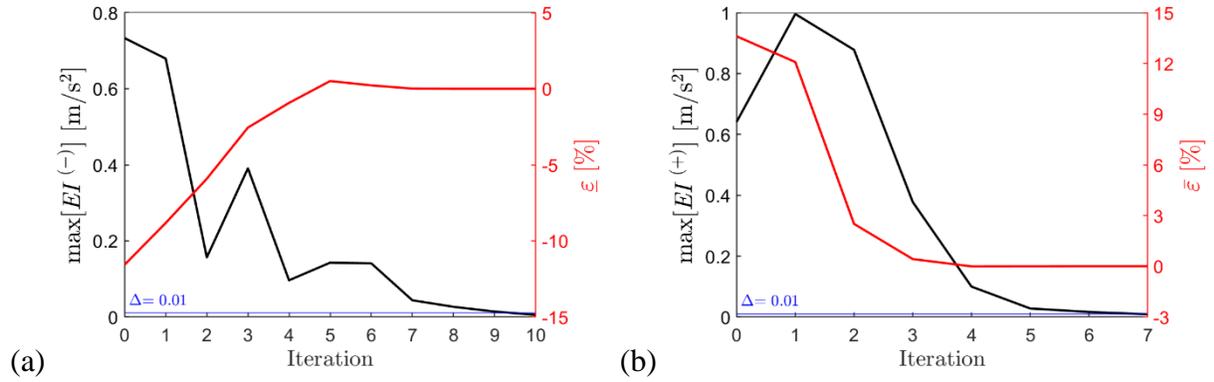

Figure 9. Variation of the maximum of the EI (black solid line) and percentage error in the response bounds estimate (red solid line) at each iteration until the relative stopping criterion $\Delta = 0.01$ m/s$^2$ (blue line) is reached for the LB (a) and UB (b) response prediction. The constant blue line corresponds to the tolerance value $\Delta$.

**CBs AF.** The stopping criterion of the CBs AFs are expressed in terms of the observed maximum and minimum of the response at each iteration (Eq. (33)). It is possible to observe that these criteria are achieved, respectively, at iterations 12 (Figure 10a) and 7 (Figure 10b), and that the LB and UB are estimated with $\underline{\varepsilon} = 0\%$ and $\bar{\varepsilon} = 0\%$. Therefore, 22 simulations were run.

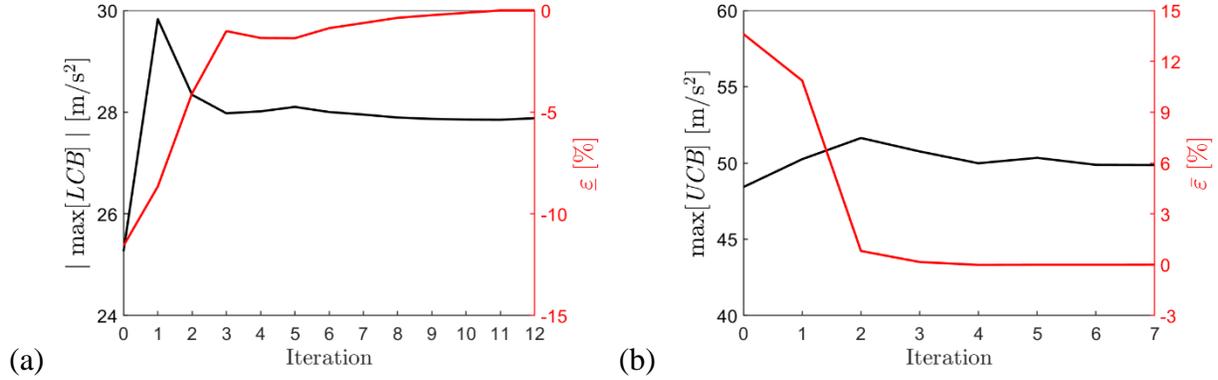

Figure 10. Variation of the absolute maximum of the CBs (black solid line) and percentage error in the response bounds estimate (red solid line) at each iteration until the relative stopping criterion are reached for the LB (a) and UB (b) response prediction.

In conclusion, for the case under investigation, the stopping criterion on AF was successfully used to reduce the number of simulations to be run without compromising the accuracy on the bound estimates when considering Approach A. Moreover, EI and CBs showed a similar performance.

### 5.1.5 Approach B with stop criterion on AF: investigation on AFs choice

**EI AF.** For the EI AF, the stop criterion is set such that $\Delta = 10^{-2}$ m/s$^2$. By inspection of Figure 11a and 11b, it is possible observe that the stopping criteria for the assessment of the LB and UB are achieved at the 7$^{th}$ iteration, and the LB and UB are estimated with percentage errors of $\underline{\varepsilon} = 0\%$ and $\bar{\varepsilon} = 0\%$, respectively. In Approach B, at each iteration 2 simulations are evaluated. Therefore, the total computational budget is of 17 simulations (3 initials, 7 for the UB and 7 for the LB). It is important to note that as the number of iterations increases, the functions $\max\left[EI^{(-)}\right]$ and $\max\left[EI^{(+)}\right]$ obtained using Approach A (Figure 9a and 9b) are different than that obtained using Approach B (Figure 11a). This is because in Approach A two separate enhanced training dataset are constructed, while in Approach B only one training dataset is built. This produces different predictive mean functions and confidence bounds, which in turn leads to the selection of different points for improving the bounds prediction.

Figure 12 shows the GP evolution for each iteration obtained by using Approach B together with the location of the next two points to be evaluated to improve the LB and UB, indicated respectively with blue and red vertical lines.

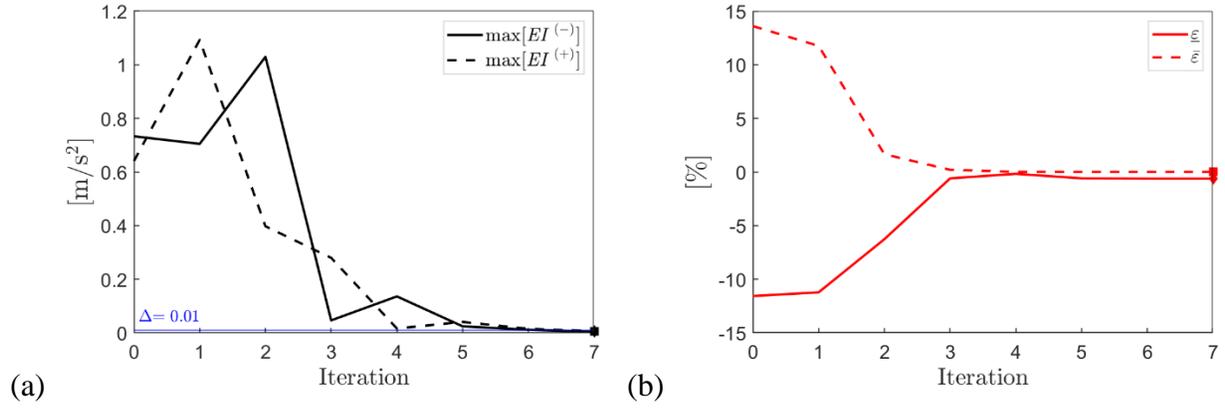

Figure 11. Approach B results for each iteration expressed in terms of: (a) $\max\left[EI^{(-)}\right]$ (solid line) and $\max\left[EI^{(+)}\right]$ (dashed line); (b) percentage errors $\underline{\varepsilon}$ (solid line) and $\bar{\varepsilon}$ (dashed line). The constant blue line corresponds to the tolerance value $\Delta$.

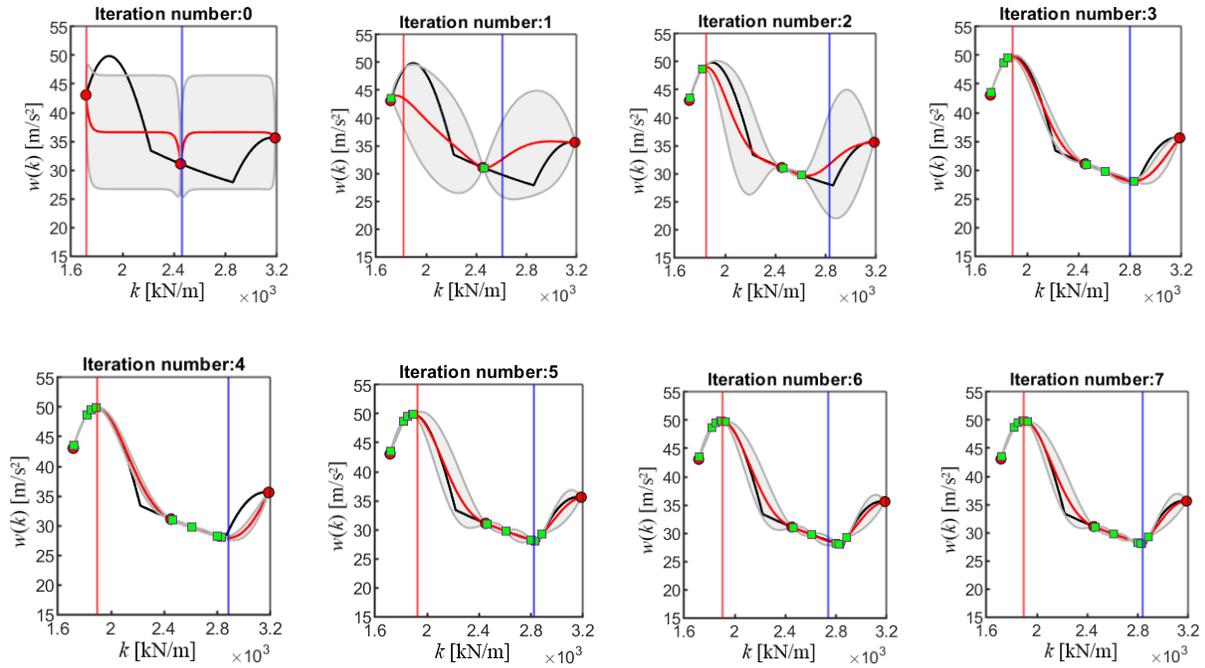

Figure 12. Evolution of the mean function (red curve) and confidence interval (shaded gray area) of the GP regression model for evaluating the UB and LB of the response (Approach B) by including the selected simulation results (circle and square markers) at iteration 1-7 when using the EI AFs. The results are compared to the SM results (black curve). The blue vertical solid lines represent the location of the next candidate point for evaluating the minimum. The red vertical solid lines represent the location of the next candidate point for evaluating the minimum.

**CBs AF.** In Figure 13a and Figure 13b it is possible to observe that the stopping criteria for the assessment of the LB and UB are achieved respectively at the 14th and the 7th iterations, with $\underline{\varepsilon} = 0\%$ and $\overline{\varepsilon} = 0\%$. Beyond the 3 initial simulations, 14 simulations were run until the 7th iteration when the stopping criterion on the LCB was reached, and only 7 simulations were run until the 14th iteration, requiring a total computational budget of 24 simulations.

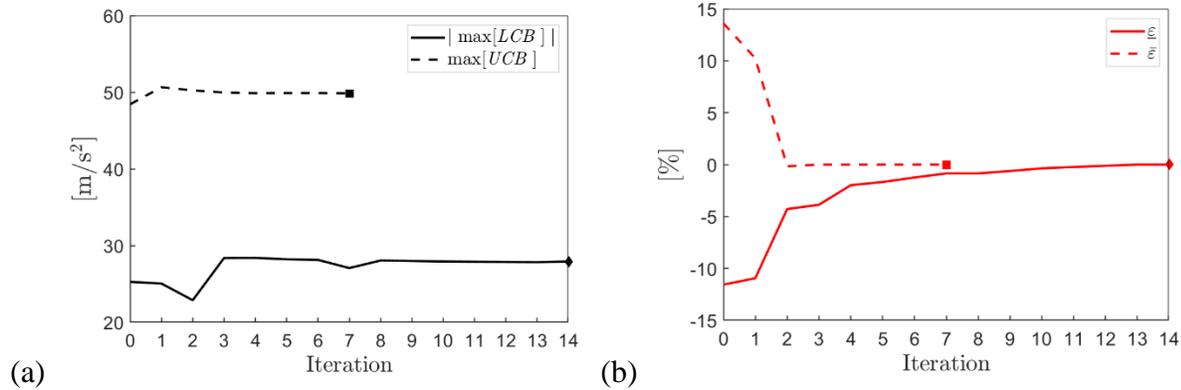

(a) (b)

Figure 13. a) $|\max[LCB]|$ (solid line) and $\max[UCB]$ (dashed line) b) percentage difference $\underline{\varepsilon}$ (solid line) and $\overline{\varepsilon}$ (dashed line) versus the number of iterations of Approach B.

### 5.1.6 Results comparison for the last iteration

The results obtained with Approach A and Approach B when considering the stopping criterion on the AF are summarised in Table 1.

|  | Sub-interval Method | Vertex Method | Approach A (EI AF) | Approach B (EI AF) | Approach A (CBs AF) | Approach B (CBs AF) |
|---|---|---|---|---|---|---|
| $\underline{w}, k\|_{\underline{w}}$ | 27.856 m/s², 2856.70 kN/m | 35.594 m/s², 3185.00 kN/m | 27.856±0 m/s², 2856.70 kN/m | 28.028±0 m/s², 2832.20 kN/m | 27.856±0 m/s², 2856.70 kN/m | 27.856±0 m/s², 2856.70 kN/m |
| $\overline{w}, k\|_{\overline{w}}$ | 49.866 m/s², 1891.40 kN/m | 43.081 m/s², 1714.00 kN/m | 49.866±0 m/s², 1896.30 kN/m | 49.866±0 m/s², 1896.30 kN/m | 49.866±0 m/s², 1891.4 kN/m | 49.866±0 m/s², 1891.4 kN/m |
| $w_{\min}, k\|_{w_{\min}}$ | - | - | 27.856 m/s², 2856.70 kN/m | 28.028 m/s², 2832.20 kN/m | 27.856 m/s², 2856.70 kN/m | 27.856 m/s², 2856.70 kN/m |
| $w_{\max}, k\|_{w_{\max}}$ | - | - | 49.866 m/s², 1896.30 kN/m | 49.866 m/s², 1896.30 kN/m | 49.866 m/s², 1891.40 kN/m | 49.866 m/s², 1891.40 kN/m |
| $\underline{\varepsilon}$ wrt SM | - | 27.78% | 0% | 0% | 0% | 0% |
| $\overline{\varepsilon}$ wrt SM | - | 13.60% | $1.706 \times 10^{-3}\%$ | 0% | 0% | 0% |
| **Total number of simulations** | 301 | 2 | 20 | 17 | 22 | 24 |

Table 1. Summary of results for first case study

It is possible to conclude that Approach A and Approach B lead to a drastic reduction of the total number of simulations to be carried out compared with the SM, while yielding very accurate prediction of the response bounds. Moreover, EI and CBs AFs perform equally well for the problem under investigation.

### 5.1.7 Effect of initial training dataset on Approach A and Approach B

Let us consider $n$ equal partitions of the interval variable leading to $n+1$ values of the interval variable at which the response has to be evaluated in order to define the initial training dataset $D_{n+1}$. Both the EI and CBs AFs are considered, and the same tolerances as before are implemented for the stopping criteria on each AF. Five cases are going to be explored when $n = [2,5,8,11,14]$. From Figure 14, it is possible to observe that the number of total simulations to be evaluated to reach a specified stopping criterion on the AFs, does not decrease as the number of elements in $D_{n+1}$ increases. For both approaches A and B, the number of total simulations is reduced only for $D_6$ when EI is used, and $D_9$ when CBs is used.

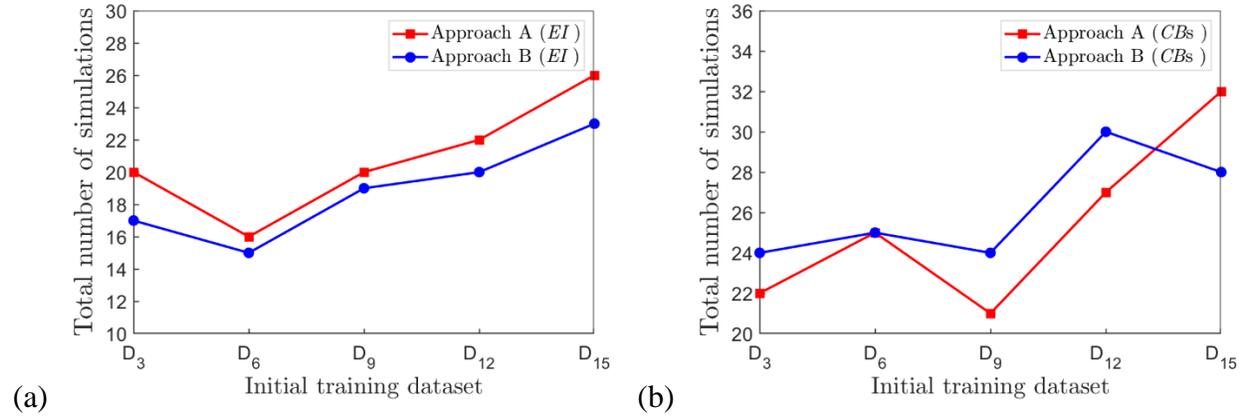

Figure 14. Total number of simulations required by Approach A and Approach B for different initial training datasets and using EI (a) and CB (b) as acquisition functions

The absolute percentage error on the bound estimates obtained at the last iteration carried out with Approach A and Approach B considering different initial training dataset, when EI and CBs AFs are considered, are shown in Figure 15 and 16, respectively. It can be observed that the percentage error is below 1% and does not decrease for increasing number of initial simulations or total number of simulations.

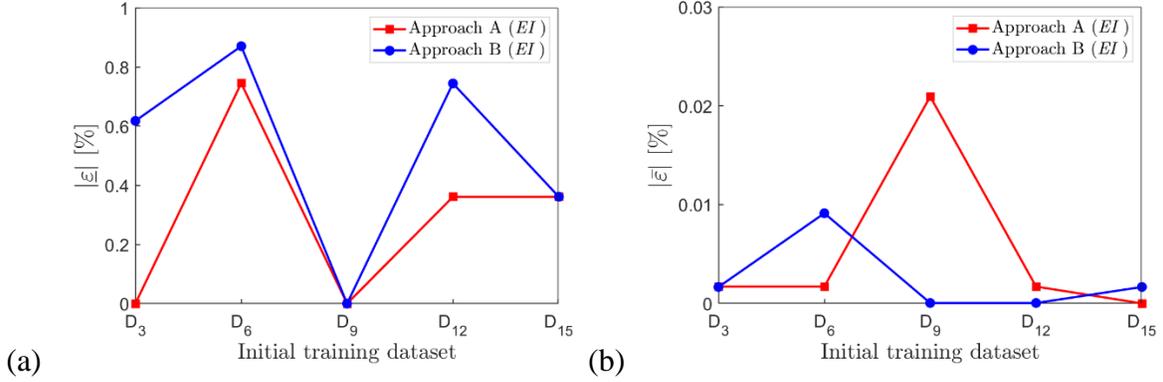

Figure 15. Absolute percentage error on the lower bound (a) $\underline{\varepsilon}[\%]$ and upper bound (b) $\overline{\varepsilon}[\%]$ obtained for Approach A (red curve) and Approach B (blue curves), considering different initial training datasets and using EI as AF.

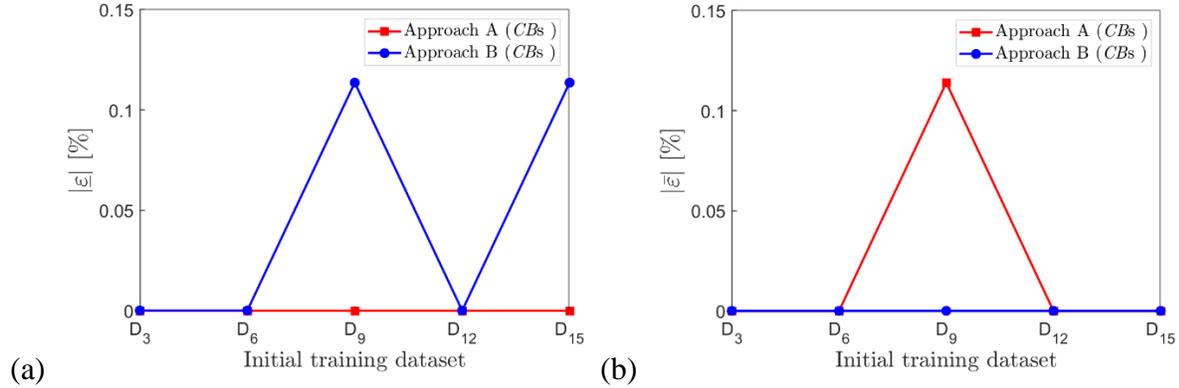

Figure 16. Absolute percentage error on the lower bound (a) $\underline{\varepsilon}[\%]$ and upper bound (b) $\overline{\varepsilon}[\%]$ obtained for Approach A (red curve) and Approach B (blue curves), considering different initial training datasets and using CBs as AF.

These results can be explained by considering that by evaluating simulations which are not specifically selected to improve the response bound estimates, the regression model of the overall response would improve globally, but this will not necessarily speed up the search of the response bounds. While in Approach A, this problem is related only to the size of the initial training dataset, in Approach B this effect is also caused by the fact that a single GP is updated to solve for both the upper and lower bounds. In Figure 17, the results expressed in terms of the intervals on the bound estimates and the minimum and maximum of the observations when considering different initial training datasets are shown. It can be observed that increasing the initial training datasets, does not necessarily lead to an improvement on the bound estimates.

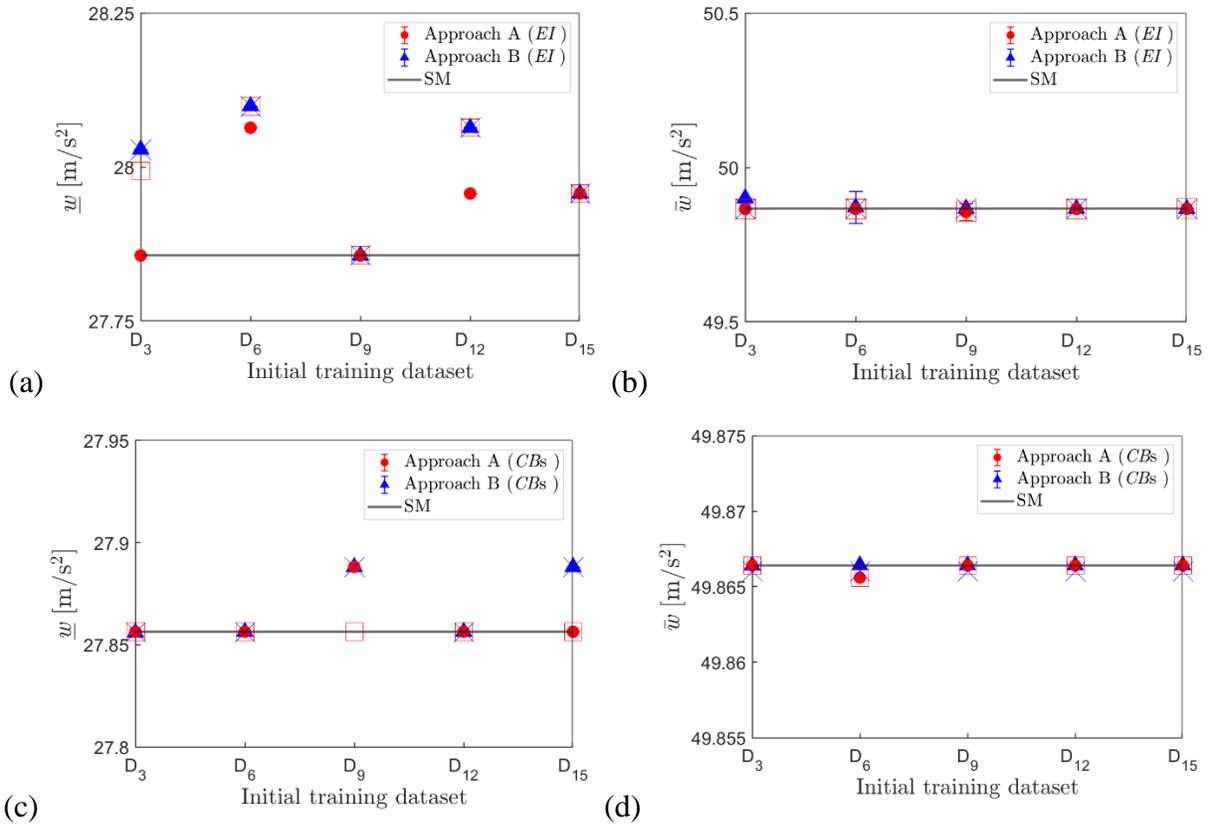

Figure 17. Response bounds obtained with Approaches A (in red) and B (in blue). EI AF for (a) lower bound and (b) upper bound estimates. CBs AF for (a) lower bound and (b) upper bound estimates. The minimum and maximum of the observations obtained when using Approach A are indicated with a red square marker, and those obtained when using Approach B are indicated with the a blue x marker. The constant black line corresponds to the exact bound.

For most of the cases investigated, the conditions defined in Eq (40) – Eq. (47) are met. Eq. (41) is not met for the cases $D_3, D_6$ when using the EI AF for evaluating the UB with Approach A. The condition expressed with Eq. (40) is not met for any of the initial training datasets investigated when using the CBs AF for evaluating the LB with Approach A. This indicates that the AFs were suggesting to explore a point of the interval varaible domain distant from the current predicted minimum/maximum of the GP. However, Eq. (42) and Eq. (43) were verified, therefore, no warning was raised. Indeed, the results reported in Figure 17 show a good accuracy in the prediction of the bounds.

### 5.2 Plate-acoustic-cavity system with four interval variables

Enclosures with one flexible wall excited by an external loading are commonly used to provide insights into a variety of structural–acoustic problems [55]. In this section, we will investigate the vibro-acoustic response of a flexible simply-supported rectangular plate coupled to a rigid-wall acoustic cavity, driven by a single unit harmonic force applied to the plate. This problem is of interest since the plate vibration will produce an acoustic pressure in the cavity, which needs to be assessed and kept below some specific thresholds to satisfy noise level requirements. The cavity contains air, and therefore the plate and cavity are weakly coupled, and the modal-interaction approach is used to evaluate the responses of interest. Let us consider the flexible simply supported rectangular aluminium plate coupled with an acoustic cavity with 5 rigid and perfectly reflecting walls, as described in [55], whose model parameters are summarised in Table 2.

|  | Width [m] | Height [m] | Thickness/ Depth [m] | Density [kg/m^3] | Plate Poisson's ratio |
|---|---|---|---|---|---|
| **Plate** | 0.5 | 0.3 | Interval variable | 2700 | 0.3 |
| **Cavity** | 0.5 | 0.3 | 1.1 | Interval variable | - |

Table 2. Plate-acoustic-cavity system model properties, as reported in [55].

The plate thickness, plate Young modulus, fluid density and fluid speed of sound are assumed to be described by interval variables and are summarised in Table 3.

|  | Lower Bound | Nominal Value | Upper bound |
|---|---|---|---|
| **Plate thickness [m]** | 0.0028 | 0.003 | 0.0032 |
| **Plate Young modulus [Pa]** | $70 \times 10^9$ | $70.95 \times 10^9$ | $71.9 \times 10^9$ |
| **Fluid Density [kg/m^3]** | 1.20 | 1.21 | 1.22 |
| **Fluid speed of sound [m/s]** | 342 | 344 | 346 |

Table 3. Interval variables for second case study (nominal values corresponding to those in [55]).

A constant unit amplitude force defined in the the frequency range 0-400 Hz is applied on the plate at $(0.10 \text{ m}, 0.0075 \text{ m}, 0.00 \text{ m})$ [55]. The plate vibration generates a variation in the Sound Pressure Level (SPL) of the acoustic cavity which is measured with a microphone at $(0.125 \text{ m}, 0.150 \text{ m}, -0.875 \text{ m})$ [55]. For the present investigation, the response variable of interest is the SPL at the microphone location calculated at 109 Hz. This frequency is close to the first resonance of the plate when the nominal values are considered. Therefore, it is anticipated that the uncertainties in the plate thickness will strongly affect the SPL at the frequency of interest.

Reference [55] provides the Ansys FE model of this system for the nominal properties indicated in Table 3 considering a frequency range up of 1-400 Hz. In particular, FLUID20 elements were used for the acoustic domain, while SHELL181 elements were used for the plate [55]. Each element size was set to 0.025 m [55]. More details on the FE model can be found in [55]. For a given set of the interval variable (e.g. considering the nominal values), this relatively simple model can take several hours to run [55]. In the same reference [55], this FE model is verified with a very fast numerical model. For speeding up the computational cost of the current investigations, the fast numerical model available in [55] is used. Damping is included in the calculation by means of the application of the correspondence principle of linear viscoelasicity [59], which allows solving the damped problem by replacing the stiffness matrix of the uncoupled plate and of the uncoupled acoustic cavity with the corresponding complex modulus $\mathbf{K}(1+i\eta)$, where $\eta$ is the loss factor and $\mathbf{K}$ is the stiffness matrix. The loss factors of the plate and of the cavity are set to 0.03.

The results obtained by applying the SM (40 subintervals for each interval variable) and the VM are shown in Table 4. Scaling of each input interval variable was carried out so that all the intervals were of a similar magnitude, and the objective function was modified consequently.

|  | Sub-interval Method | Vertex Method |
| --- | --- | --- |
| **Number of simulations** | 2,825,761 | 16 |
| **Response LB [Pa]** | 1.5382 | 1.5382 |
| $\underline{\varepsilon}$ **wrt SM** | - | 0% |
| **Coordinates LB [mm, Pa, kg/m^3, m/s]** | $[320\times10^{-2}, 719.000\times10^{8}, 120.000\times10^{-2}, 346.0]$ | $[320\times10^{-2}, 719.000\times10^{8}, 120.000\times10^{-2}, 346.0]$ |
| **Response UB [Pa]** | 11.4765 | 4.9579 |
| $\overline{\varepsilon}$ **wrt SM** | - | 56.80% |
| **Coordinates UB [mm, Pa, kg/m^3, m/s]** | $[290\times10^{-2}, 719.000\times10^{8}, 122.000\times10^{-2}, 346.0]$ | $[280\times10^{-2}, 719.000\times10^{8}, 122.000\times10^{-2}, 346.0]$ |

Table 4. Results obtained with Sub-Interval Method and Vertex Method for the plate-acoustic-cavity system.

It is possible to observe that the LB occurs at one of the input bounds, and thus it is correctly evaluated with the VM, while the estimate of the UB obtained with the VM is 57% lower than that obtained with the SM. This is because of the resonance effect. If is also worth noting that the results obtained at the nominal values $\left[300\times10^{-2}\,\text{mm}, 709.500\times10^{8}\,\text{Pa}, 121.000\times10^{-2}\,\text{kg/m}^{3}, 344.0\,\text{m/s}\right]$ yield a SPL of 5.7191 Pa.

### 5.2.1 Approach A and B with fixed computational budget and EI as AF

Approach A and Approach B are applied with EI as the AF. A total computational budget of 100 simulations is considered in addition to the simulations evaluated for building the initial training dataset. In particular, four initial training datasets were considered by running 8, 9, 16 and 25 simulations. These simulations were selected by applying the Taguchi method, and the set of input variables considered for each initial training dataset are specified in appendix. The 100 simulations were divided equally between LB and UB, unless the stopping criterion on the AF was met for one bound. For this numerical application only at $D_{16}$ the stopping criterion on the EI AF at the last iteration was met, as shown in Figure 18.

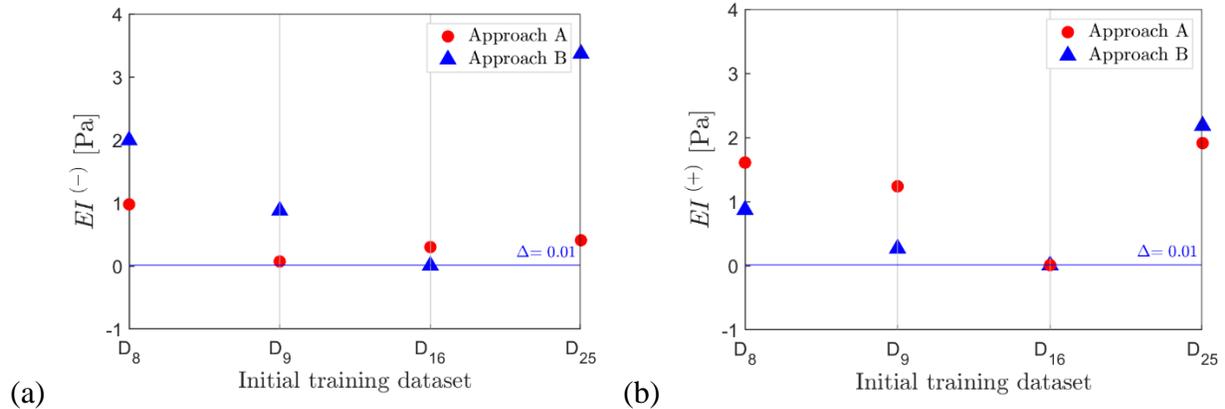

Figure 18. Values of the Expected Improvement (EI) acquisition function obtained with Approach A (in red) and Approach B (in blue) at four initial training datasets when estimating the Lower Bound (a) and Upper Bound (b). The minimum and maximum of the EI obtained when using Approach A are indicated with a red circle marker, and those obtained when using Approach B are indicated with the a blue triangle marker. The constant blue line corresponds to the tolerance value $\Delta$.

The number of simulations considered for each analysis is summarised in Table 5. Only for Approach B with $D_{16}$ the stopping criterion on the EI AF is met for both the LB and UB.

| Initial Training dataset | Approach A | | Approach B | |
|---|---|---|---|---|
| | # of simulations for meeting stop criteria for LB | # of simulations for meeting stop criteria for UB | # of simulations for meeting stop criteria for LB | # of simulations for meeting stop criteria for UB |
| **D8** | Not met, 50 | Not met, 50 | Not met, 50 | Not met, 50 |
| **D9** | Not met, 50 | Not met, 50 | Not met, 50 | Not met, 50 |
| **D16** | Not met, 90 | Met, 10 | Met, 13 | Met, 69 |
| **D25** | Not met, 50 | Not met, 50 | Not met, 50 | Not met, 50 |

Table 5. Number of simulations used in Approach A and Approach B for four initial training datasets.

The intervals on the bound estimates (minimum and maximum of the mean function) and also the minimum and maximum of the observations for four initial training datasets are shown in Figure 19. It can be observed that, as expected, the bound estimates are consistently over-predicting the maximum of the deterministic observations, and consistently under-predicting the minimum of the deterministic observations. Therefore, the bound estimates provide more conservative results. Moreover, as for the previous numerical applications, increasing the elements within initial training datasets, does not necessarily lead to an improvement on the bound estimates. It can also be observed that Approach B might largely under-predict the LB. However, a different pattern in observed for the UB. The most interesting results are obtained for $D_9$, where the maximum of the mean function obtained with both Approach A and Approach B is under-predicting the upper bound, and the confidence interval is misleadingly suggesting to trust those estimates. As a result, the condition expressed in Eq. (47) would be met. For Approach B with $D_{16}$, a relatively good agreement between the prediction and the exact bounds can be observerd in Figure 19.

A closer look at the absolute percentage error on the bound estimates obtained at the last iteration of Approach A and Approach B when compared to the SM results is shown in Figure 20. It can be observed that the percentage error can be very large, especially for the lower bound, and that it does not necessarily decrease for increasing number of initial simulations or total number of simulations. The results obtained for Approach B with $D_{16}$, have satisfied the stopping criterion on EI, however they display some percentage errors on the bound estimates.

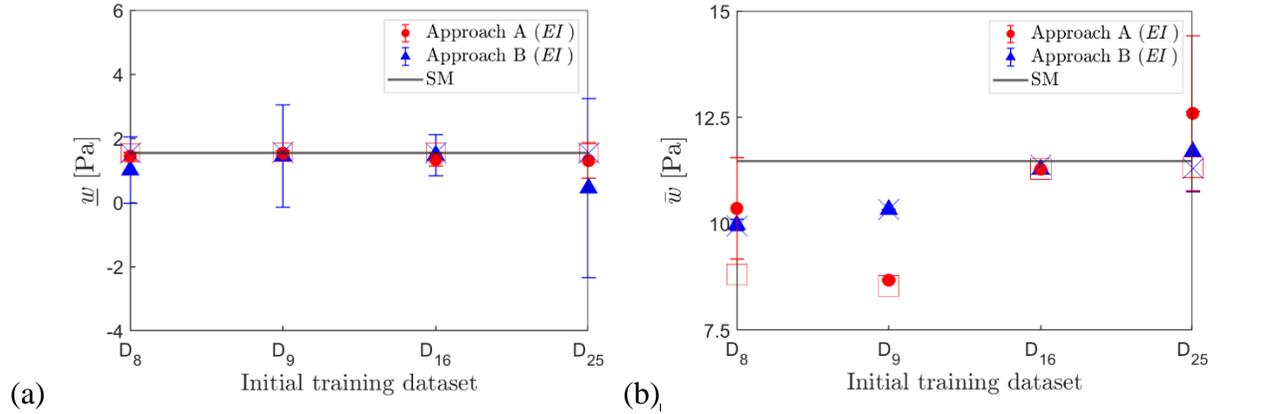

(a)　(b)

Figure 19. Lower Bounds (a) and Upper Bounds (b) estimates obtained with Approach A (in red) and Approach B (in blue) for different initial training dataset and considering EI AF. The minimum and maximum of the observations obtained when using Approach A are indicated with a red square marker, and those obtained when using Approach B are indicated with the a blue x marker. The constant black line corresponds to the exact bound.

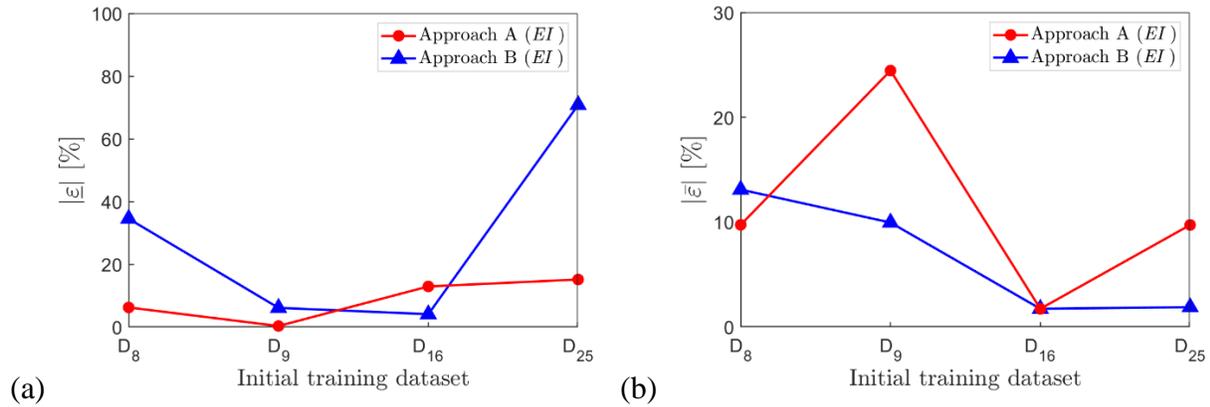

(a)　(b)

Figure 20. Absolute percentage error on the lower bound (a) $\underline{\varepsilon}[\%]$ and upper bound (b) $\overline{\varepsilon}[\%]$ obtained for Approach A (red curve) and Approach B (blue curves), considering different initial training datasets and using EI as AF.

Since this information on the percentage error is considered not available (that is, the full SM analysis was not carried out), the conditions described in Section 2.8 were assessed at the last iteration for each interval variable (plate thickness, plate Young modulus, fluid density, fluid speed of sound). The results are summarized in Tables 6 and 7, where the letter Y is used to indicate when the condition was met; otherwise, the letter N was used.

|  | **Approach A (EI)** [Eq. (40) ], Eq. (42) [Eq. (44)], Eq. (46) | **Approach B (EI)** [Eq. (40) ], Eq. (42) [Eq. (44)], Eq. (46) |
|---|---|---|
| **D8** | [N N Y N], N <br> [Y Y Y N], N | [N N Y N], **Y** <br> [**Y** N Y N], N |
| **D9** | [**Y** Y N Y], N <br> [**Y** Y N N], **Y** | [**Y** N N N], **Y** <br> [**Y** Y N N], N |
| **D16** | [**Y** N N N], N <br> [**Y Y** N N], N | [N N N N], **Y** <br> [**Y Y** N **Y**], **Y** |
| **D25** | [N N Y N], N <br> [Y N N N], N | [N N Y N], **Y** <br> [Y N N N], N |

Table 6. Assessment of the conditions for the LB response when using EI for Approach A and for Approach B

|  | **Approach A (EI)** [Eq. (41) ], Eq. (43) [Eq. (45)], , Eq. (47) | **Approach B (EI)** [Eq. (41) ], Eq. (43) [Eq. (45)], Eq. (47) |
|---|---|---|
| **D8** | [**Y Y Y Y**], N <br> [Y Y N Y], N | [N N N Y], **Y** <br> [Y Y N Y],**Y** |
| **D9** | [N N N N] N <br> [Y Y N Y], **Y** | [**Y** N N N], **Y** <br> [**Y Y Y Y**], **Y** |
| **D16** | [N N **Y** N], **Y** <br> [**Y Y Y Y**], **Y** | [N **Y Y** N], **Y** <br> [**Y Y** N N],**Y** |
| **D25** | [Y N Y N], N <br> [Y N N N], N | [Y N N N] , **Y** <br> [Y N N N], **Y** |

Table 7. Assessment of the conditions for the UB response when using EI for Approach A and for Approach B

As anticipated, it is possible to observe that there was no instance in which all the conditions (Eq, (40) –Eq. (47)) were simultaneously met. It is worth noting that when conditions expressed in Eq. (44) and (45) are met, the points at which the predicted bounds and those yielded by the deterministic simulations are very close to each other, or coincident. While when the conditions expressed in Eq. (46) and (47) are met, the magnitude of the responses at those points are very close or coincident. However, this does not ensure that an accurate estimate of the maximum has been obtained – as shown in Figure 20b, for the case of $D_9$. Regarding the results obtained with $D_{16}$ when using Approach B: although the stopping criteria on the EI AF for both the LB and UB were met, there are small percentage errors in the magnited of the predicted bounds. It is also worth noting that for the UB at $D_{16}$, Approach A and Approach B show a very similar percentage error (Figure 20b), but the two approaches yield different bound estimates. This can be also observed in Tables 8 and 9, where the position of the identified LB and UB obtained with Approaches A and B are reported. From these tables, it is possible to observe errors smaller than 3% for all the input interval variables, which nonetheless lead to large variations in the response bounds.

|  | **Approach A, LB** | **Percentage error, LB** | **Approach A, UB** | **Percentage error, UB** |
|---|---|---|---|---|
| **D8** | $[320\times10^{-2}, 719.000\times10^{8}, 120.000\times10^{-2}, 342.2]$ | [0%, 0%, 0%, 1.1%] | $[290\times10^{-2}, 700.950\times10^{8}, 121.5\times10^{-2}, 346.0]$ | [0% 2.5% 0.4% 0%] |
| **D9** | $[320\times10^{-2}, 719.000\times10^{8}, 121.800\times10^{-2}, 342.2]$ | [0%, 0%, -1.6%, 1.1%] | $[290\times10^{-2}, 700.000\times10^{8}, 122.0\times10^{-2}, 344.9]$ | [0% 2.6% 0% 0.3%] |
| **D16** | $[320\times10^{-2}, 719.000\times10^{8}, 121.300\times10^{-2}, 343.6]$ | [0%, 0%, -1.1%, 0.7%] | $[290\times10^{-2}, 719.000\times10^{8}, 120.0\times10^{-2}, 343.3]$ | [0% 0% 1.6% 0.8%] |
| **D25** | $[320\times10^{-2}, 701.425\times10^{8}, 120.000\times10^{-2}, 343.8]$ | [0%, 2.4%, 0%, 0.6%] | $[290\times10^{-2}, 709.500\times10^{8}, 120.0\times10^{-2}, 345.7]$ | [0% 1.3% 1.6% 0.1%] |

Table 8. Error on LB and UB locations for four initial training datasets (Approach A, EI)

|  | **Approach B, LB** | **Percentage error, LB** | **Approach B, UB** | **Percentage error, UB** |
|---|---|---|---|---|
| **D8** | $[320\times10^{-2}, 701.900\times10^{8}, 120.000\times10^{-2}, 342.6]$ | [0% 2.4% 0% 1.0%] | $[290\times10^{-2}, 708.075\times10^{8}, 129.500\times10^{-2}, 343.7]$ | [0% 1.5% 0% 0.7%] |
| **D9** | $[320\times10^{-2}, 719.000\times10^{8}, 121.300\times10^{-2}, 342.1]$ | [0% 0% -1.1% 1.1%] | $[290\times10^{-2}, 709.500\times10^{8}, 120.700\times10^{-2}, 346]$ | [0% 1.3% 1.1% 0%] |
| **D16** | $[320\times10^{-2}, 719.000\times10^{8}, 121.800\times10^{-2}, 343.6]$ | [0% 0% -1.5% 0.6%] | $[290\times10^{-2}, 719.000\times10^{8}, 120.000\times10^{-2}, 343.3]$ | [0% 0% 1.6% 0.8%] |
| **D25** | $[320\times10^{-2}, 709.500\times10^{8}, 120.000\times10^{-2}, 344.6]$ | [0% 1.3% 0% 0.4%] | $[290\times10^{-2}, 709.500\times10^{8}, 122.000\times10^{-2}, 345.7]$ | [0% 1.3% 0% 0.1%] |

Table 9. Error on LB and UB locations for four initial training datasets (Approach B, EI)

### 5.2.2 Approach A and Approach B with fixed computational budget, CBs AF

Also for the case of the CBs as AF, 100 simulations in addition to each initial training dataset were considered. The stopping criteria on the AF for the LB has been met for both approaches, while the stopping criterion on the AF was not met for the UB, as shown in Figure 21. The total number of simulations used in Approach A and Approach B are reported in Table 10.

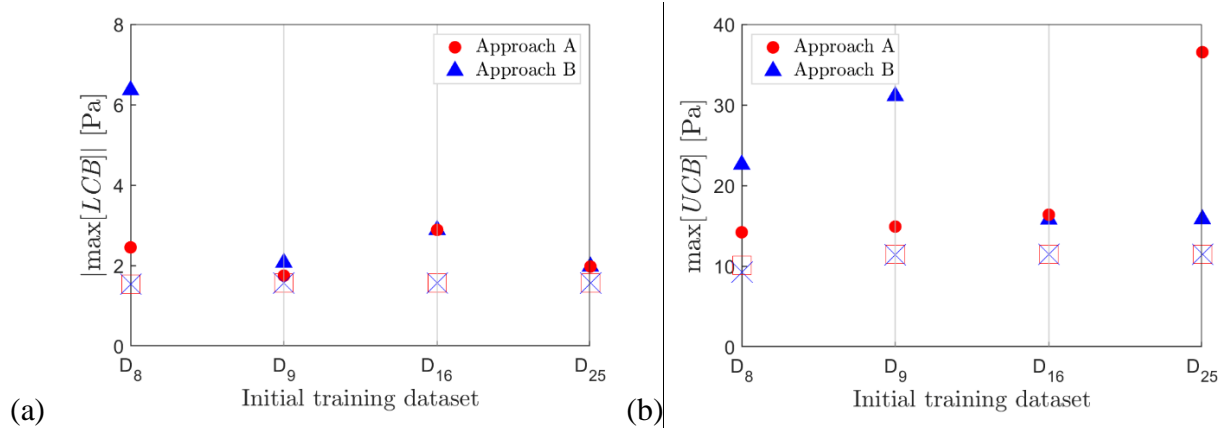

(a)  (b)

Figure 21. Absolute maximum of the LCBs (a) and of the UCBs (b) obtained with Approach A (in red) and Approach B (in blue) for four initial training datasets. The minimum and maximum of the AF obtained at the last iteration are indicated with a red circle marker (Approach A) and blue triangle marker (Approach B). The minimum and maximum of the observations obtained are indicated with a red square marker (Approach A) and blue x marker (Approach B).

| Initial Training dataset | Approach A | | Approach B | |
|---|---|---|---|---|
| | # of simulations for meeting stop criteria for LB | # of simulations for meeting stop criteria for UB | # of simulations for meeting stop criteria for LB | # of simulations for meeting stop criteria for UB |
| D8 | Met, 4 | Not met, 96 | Met, 21 | Not met, 79 |
| D9 | Met, 13 | Not met, 87 | Met, 19 | Not Met, 81 |
| D16 | Met, 1 | Not met, 99 | Met, 1 | Not met, 99 |
| D25 | Met, 1 | Not met, 99 | Met, 1 | Not met, 99 |

Table 10. Number of simulations used in Approach A and B for four initial training datasets.

The results expressed in terms of intervals on the bound estimates and in terms of the minimum and maximum of the observations for the four initial training datasets are shown in Figure 22. The corresponding absolute percentage error on each bound estimate obtained at the last iteration of Approach A and Approach B are shown in Figure 23. One of the key results shown in Figure 23(a) is that even if the conditions on the AF are met for all the LB analyses, there are errors in the response estimates. Moreover, also in this case increasing the elements within the initial training datasets, does not necessarily lead to an improvement on the bound estimates. Overall, percentage errors below 20% were observed for both Approach A and Approach B, therefore, for the given computational budget and the same set of initial training dataset, the CBs AF performs better than the EI AF.

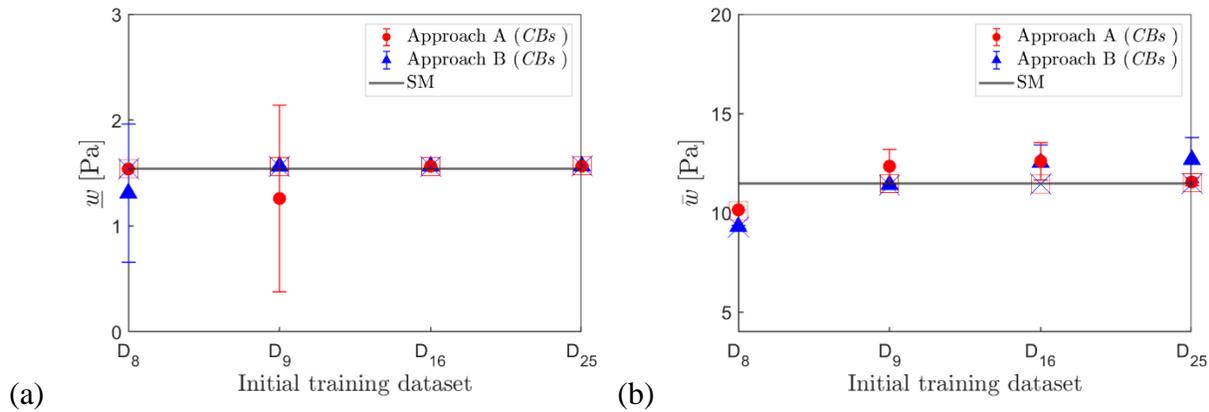

(a) (b)

Figure 22. Lower Bounds (a) and Upper Bounds (b) estimates obtained with Approach A (in red) and Approach B (in blue) for different initial training dataset and considering CBs AF. The minimum and maximum of the observations obtained when using Approach A are indicated with a red square marker, and those obtained when using Approach B are indicated with the a blue x marker. The constant black line corresponds to the exact bound.

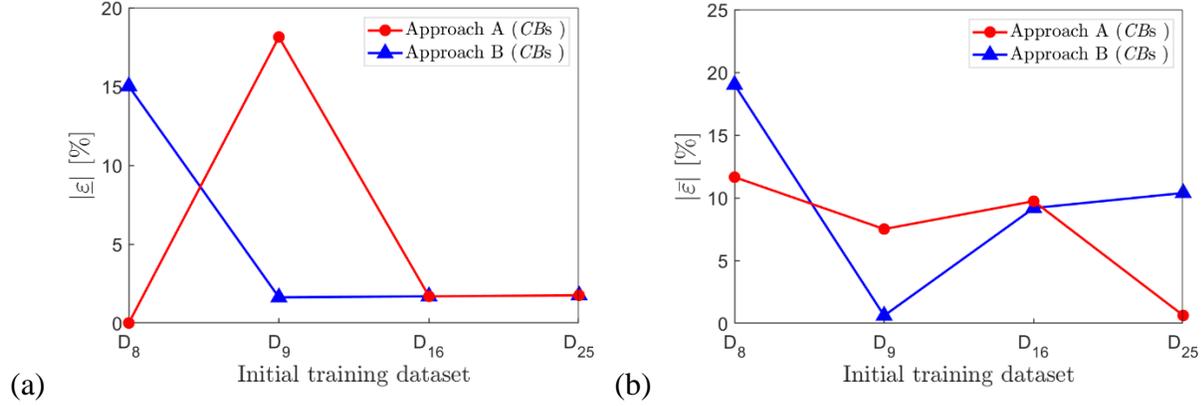

Figure 23. Absolute percentage error on the lower bound (a) $\underline{\varepsilon}[\%]$ and upper bound (b) $\bar{\varepsilon}[\%]$ obtained for Approach A (red curve) and Approach B (blue curves), considering different initial training datasets and using CBs as AF.

The conditions described in Section 2.9 were considered also in this case. The results are summarized in Table 11 and 12, where the letter Y is used to indicate when the condition was met, otherwise the letter N was used.

|  | Approach A (CBs) [Eq. (40) ], Eq. (42) [Eq. (44)], Eq. (46) | Approach B (CBs) [Eq. (40) ], Eq. (42) [Eq. (44)], Eq. (46) |
|---|---|---|
| **D8** | [N Y N N], N<br>[**Y Y Y Y**], **Y** | [N N Y N], **Y**<br>[Y Y Y N], N |
| **D9** | [Y Y N N], N<br>[Y Y N N], N | [N Y N N], N<br>[**Y Y Y Y**], **Y** |
| **D16** | [N Y N N], N<br>[**Y Y Y Y**], **Y** | [N Y N N], N<br>[**Y Y Y Y**], **Y** |
| **D25** | [N Y N Y], N<br>[**Y Y Y Y**], **Y** | [N Y N Y], N<br>[**Y Y Y Y**], **Y** |

Table 11. Conditions for LB Approach A and Approach B when using CBs

|  | Approach A (CBs) [Eq. (41) ], Eq. (43) [Eq. (45)], Eq. (47) | Approach B (CBs) [Eq. (41) ], Eq. (43) [Eq. (45)], Eq. (47) |
|---|---|---|
| **D8** | [Y N N N], N<br>[**Y Y Y Y**], **Y** | [N N N N], **Y**<br>[Y Y N N], **Y** |
| **D9** | [Y Y N N], N<br>[Y Y N N], N | [N N N N], **Y**<br>[Y Y N Y], **Y** |
| **D16** | [Y Y N N], N<br>[Y Y N N], N | [Y Y N N], **Y**<br>[Y Y N N], N |
| **D25** | [N N N N], N<br>[Y Y Y N], **Y** | [N N N N], **Y**<br>[Y Y N N], N |

Table 12. Conditions for UB Approach A and Approach B when using CBs

It is worth noting that even when considering the LB obtained with Approach A with $D_8$, the conditions in Eq. (40) and Eq. (42) are not met. As anticipated, for Approach A, this can occur since the AF would suggest to explore other regions of the objective function which are currently unexplored. The errors on the location of the LB and UB are reported in Table 13 and 14, where it is possible to observe an error smaller than 2.1% for all the input interval variables.

|  | **Approach A, LB** | **Percentage error, LB** | **Approach A, UB** | **Percentage error, UB** |
|---|---|---|---|---|
| **D8** | $[320\times10^{-2}, 719.000\times10^8, 120.000\times10^{-2}, 346]$ | [0% 0% 0% 0%] | $[290\times10^{-2}, 708.550\times10^8, 120.0\times10^{-2}, 346.0]$ | [0% 1.4% 1.6% 0%] |
| **D9** | $[320\times10^{-2}, 719.000\times10^8, 121.400\times10^{-2}, 342]$ | [0% 0% -1.2% 1.2%] | $[290\times10^{-2}, 719.550\times10^8, 121.6\times10^{-2}, 345.4]$ | [0% 0% 0.3% 0.2%] |
| **D16** | $[320\times10^{-2}, 719.000\times10^8, 121.300\times10^{-2}, 343.3]$ | [0% 0% -1.1% 0.8%] | $[290\times10^{-2}, 719.000\times10^8, 121.6\times10^{-2}, 344.1]$ | [0% 0% 0.3% 0.5%] |
| **D25** | $[320\times10^{-2}, 719.000\times10^8, 121.500\times10^{-2}, 344.0]$ | [0% 0% -1.2% 0.6%] | $[290\times10^{-2}, 719.000\times10^8, 122.0\times10^{-2}, 344.1]$ | [0 % 0% 0% 0.5%] |

Table 13. Error on LB and UB locations for four initial training datasets (Approach A, CBs)

|  | **Approach B, LB** | **Percentage error, LB** | **Approach B, UB** | **Percentage error, UB** |
|---|---|---|---|---|
| **D8** | $[320\times10^{-2}, 719.000\times10^8, 120.000\times10^{-2}, 342.4]$ | [0% 0% 0% 1.0%] | $[290\times10^{-2}, 703.800\times10^8, 122.0\times10^{-2}, 343.6]$ | [0% 2.1% 0% 0.7%] |
| **D9** | $[320\times10^{-2}, 719.000\times10^8, 121.000\times10^{-2}, 342]$ | [0% 0% -0.8% 1.2%] | $[290\times10^{-2}, 719.000\times10^8, 122.0\times10^{-2}, 342.1]$ | [0% 0% 0% 1.13%] |
| **D16** | $[320\times10^{-2}, 719.000\times10^8, 121.300\times10^{-2}, 343.3]$ | [0% 0% -1.1% 0.8%] | $[290\times10^{-2}, 719.000\times10^8, 121.6\times10^{-2}, 344.1]$ | [0% 0% 0.3% 0.6%] |
| **D25** | $[320\times10^{-2}, 719.000\times10^8, 121.500\times10^{-2}, 344.0]$ | [0% 0% -1.2% 0.6%] | $[290\times10^{-2}, 719.000\times10^8, 120.3\times10^{-2}, 345.6]$ | [0% 0% 1.4% 0.1%] |

Table 14. Error on LB and UB locations for four initial training datasets (Approach B, CBs)

### 5.2.3 Investigation on the AF stopping criteria on CBs

The results obtained when considering $D_8$ and the stopping criterion on the CBs AF are reported in Table 15. The number of simulations needed for both Approach A and Approach B is much smaller than SM, with Approach B requiring less simulations. It is important to note that the largest error is encountered when estimating the LB with Approach B which is underpredicted by 15%. This is caused by an error of the order of 1% in the location of the bound. Moreover, it is possible to note that only for this estimate there is a confidence interval on the prediction, that is: the LB is predicted at a point for which a simulation has not been evaluated. These results would immediately flag the need for further model evaluations to be carried out to update the GP regression model and increase the confidence in that bound estimate. It is also worth noting that since $m(\mathbf{b}_{w_{\min}})$ is obtained at the LB of the response, few additional simulations might be needed.

Overall, the proposed conditions are not fully met for any of the cases shown. These suggest running further simulations, especially for the LB obtained with Approach B. Moreover, it would raise a flag for the LB obtained with Approach A, given that this has been obtained by exploring very few points of the objective function. Considering these extra simulations would have led to an improvement in the response estimates, and more confidence in the obtained estimates of the bounds.

|  | Approach A, LB | Approach B, LB | Approach A, UB | Approach B, UB |
|---|---|---|---|---|
| $m(\mathbf{b}_{LB})$ or $m(\mathbf{b}_{UB})$ [Pa] | 1.538 | 1.307 | 11.300 | 11.458 |
| $\sigma^2(\mathbf{b}_{LB})$ or $\sigma^2(\mathbf{b}_{UB})$ [PA] | 0 | 0.813 | 0 | 0 |
| $m(\mathbf{b}_{w_{min}})$ or $m(\mathbf{b}_{w_{max}})$ [PA] | 1.538 | 1.538 | 11.300 | 11.458 |
| $\varepsilon$ wrt SM in terms of bounds amplitude | 0% | 15.020% | 1.542% | 0.165% |
| $\varepsilon$ wrt SM in terms of bounds position | [0%, 0%, 0%, 0%] | [0%, 0%, 0%, 1.0%] | [0%, 0.4%, 0%, 0%] | [0%, 0.006%, 0%, 0%] |
| Conditions (Eq. (40)-Eq. (47)) | [N Y N N], N [Y Y Y Y], Y | [N N Y N], Y, [Y Y Y N], N | [Y Y N Y], Y [Y Y Y Y], Y | [Y Y N Y], Y, [Y Y Y Y], Y |
| Number of simulations | 8+4 | 8+21 | 8+577 | 421 (in addition to 8+21) |

Table 15. Results obtained for initial training dataset D8 using the stopping criteria on the CBs.

## 6. Discussion

The results presented in Section 5 indicate that:

i. The proposed approaches led to a drastic reduction of the number of simulations to be carried out in order to obtain accurate response bound estimates compared to the Sub-Interval method.

ii. Increasing the number of simulations for the initial training dataset does not necessarily lead to a reduction on the errors of the estimates of the response bounds when considering a fixed computational budget, nor would necessarily reduce the total number of simulations to be carried out to meet the stopping criteria on the AF. This is because these simulations are chosen a-priori when no knowledge about the relationships between the input and output interval variables is available. Therefore, these initial simulations would potentially improve the initial GP regression model overall, but not necessarily greatly help with the selection of the points needed to improve the prediction of the response bounds. As a result, it is recommended to consider a limited number of simulations for the initial training dataset, using for example the Taguchi method or other Design of Experiments techniques, and let the algorithm automatically select the points to be included in the enhanced training dataset using an AF that has a good balance between exploration and exploitation.

iii. The choice of the AF can have a large impact on the total number of simulations to be carried out and on the accuracy of the bound estimates, even when many simulations are included in the initial training dataset. As expected, it was shown that the Probability of Improvement should be avoided because it would favor exploitation rather than exploration. The Expected Improvement (EI) and the Confidence Bounds (CBs) Acquisition Functions have a good balance between exploitation and exploration and performed equally well. Interestingly, the CBs outperformed the EI in most of the cases investigated. Therefore, the CBs should be preferred.

iv. Estimating the response bounds considering two separate GPs (Approach A) versus a single GP (Approach B) produced different results. This is mainly because Approach A yields two separate posterior mean functions, while one posterior mean function is evaluated in Approach B. Moreover, with Approach B, the simulations included in the enhanced training dataset are not selected to improve one of the response bound estimates only. Therefore, the training dataset at each iteration enhanced the overall regression model by exploring regions of the objective function which would not have been explored otherwise, while simultaneously selecting points which would improve either the upper bound or the lower bound of the response. This reduced the possibility of the optimizer getting stuck in a local optimum point, and in some instances, it led to a reduction of the number of simulations needed to meet the criteria on the chosen AF. It can be concluded that Approach B is more robust.

v. Expressing the interval analysis results in terms of the maximum/minimum of the posterior mean function of the response function (once the stopping criteria have been met), instead of the minimum maximum of the observations, led to over-predicting the UB and under-predicting the LB with respect to the current observations available. Moreover, in the last numerical application, it was shown how the confidence bounds at those point estimates were indeed helpful in highlighting inaccuracies in the bound predictions. Large confidence intervals on the response bound estimates reflect the limited number of observations around that bound. Therefore, the larger the confidence intervals are, the lower is the confidence on the response bounds estimates obtained. Consequently, the confidence intervals can be used to justify the need of performing additional simulations or to communicate effectively the level of lack of knowledge. However, having a very small confidence interval around the bounds estimates after meeting the stopping criteria (even those on the AF) does not ensure that the correct bounds have been evaluated (as shown also in the second numerical application).

vi. Using the stopping criterion in terms of the AF is not enough for ensuring accurate estimates of the response bounds. It was shown that the conditions based on the two metrics proposed were able to flag the need for further simulations to increase the accuracy on the response bounds.

## 7. Conclusions

Two non-intrusive uncertainty propagation approaches have been proposed for the analysis of a generic engineering system subject to interval uncertainties. One of the main advantages of the proposed approaches is interpretability, since the selection of the next combinations of interval variables to be investigated is justified in terms of the trade-off between exploration and exploitation considering a probabilistic model that embeds both user knowledge and noise-free observations obtained with a physics-based model. Another advantage relates to the fact that the results are explainable and can be used to justify the decision of running additional simulations. Moreover, the bounds predicted with the proposed approach would over-predict the minimum and under-predict the true maximum of the response with respect to the current observations available. Indeed, this is an advantage in engineering applications, where a limited number of simulations can be investigated.

These approaches build a Gaussian Process (GP) of the response variable to account for the lack of knowledge caused by having evaluated the response function only for some selected combinations of the interval variables which might not necessarily correspond to the response bounds. The GP is then used in combination with the Acquisition Functions (AF) to iteratively identify the simulations that can potentially maximize the improvement on the response bound estimates. The iterations are stopped based on a computational budget criterion or a criterion on the value of the AF. These approaches enable to run a very limited number of simulations compared to alternative interval propagation approaches, moreover, they naturally embed information about the lack of knowledge related to the limited number of simulations explored. The results obtained with these approaches are estimates on the response bounds which correspond to the minimum/maximum of the mean function of the posterior distribution, together with information on the confidence on those estimates expressed in terms of confidence bounds. While Approach A addresses the evaluation of the upper and lower response bounds separately, Approach B iteratively constructs a GP model that is used to investigate both bounds. Consequently, for a fixed number of iterations, the predicted bounds obtained with the two approaches would be different. To verify that the predicted bounds are satisfactory, two metrics were proposed. One considers the distance between the next point selected by the AF and the current bound estimate. The other considers the distance between the optimum of the observations and the corresponding predicted bound. These metrics allowed the definition of some conditions that can be used to assess if additional simulations should be carried out or otherwise. The proposed approaches were applied to investigate the bounds on the vibration response of a single degree of freedom with one interval uncertainty and of the acoustic response of a plate-acoustic-cavity system with four interval uncertainties, showing excellent results.

The proposed approaches are generally applicable to the analysis of engineering systems subject to interval uncertainties for which a large-scale-model is available. However, it is well-known that GP approximations are best suited for dealing with a continuous domain with a small number of dimensions, usually less than 20 [31]. This is because, the larger the dimension of the continuous domain, the more training points are needed to fully explore the input-output relation. As a result, this might restrict its applicability to a certain number of uncertainties. Current work is focusing on investigating strategies for addressing this dimensionality issue.

**References**

[1] C. Soize, Uncertainty Quantification - An Accelerated Course with Advanced Applications in Computational Engineering, Springer, 2017.

[2] M. Wright, R. Weaver, New Directions in Linear Acoustics and Vibration. Cambridge University Press, 2010

[3] N. Metropolis, S. Ulam, The Monte Carlo method, J. Am. Stat. Assoc. 44 (1949) 335-341. https://doi.org/10.1080/01621459.1949.10483310.

[4] R.S. Langley, Unified Approach to Probabilistic and Possibilistic Analysis of Uncertain Systems, J. Eng. Mech. 126(11) (2000) 1163-1172. https://doi.org/10.1061/(ASCE)0733-9399(2000)126:11(1163).

[5] M. Faes, D. Moens, Recent Trends in the Modeling and Quantification of Non-probabilistic Uncertainty, Arch Computat Methods Eng 27 (2020) 633-671. https://doi.org/10.1007/s11831-019-09327-x.

[6] S.S. Rao, L. Berke, Analysis of Uncertain Structural Systems Using Interval Analysis, AIAA 34 (5) (1997) 727-735. https://doi.org/10.2514/2.164.

[7] Y. Ben-Haim, I. Elishakoff, Convex Models of Uncertainty in Applied Mechanics, Elsevier Science, Amsterdam, 1990.

[8] L.A. Zadeh, Fuzzy sets as a basis for a theory of possibility, Fuzzy Sets Syst. 100 (1999) 9-34. https://doi.org/10.1016/S0165-0114(99)80004-9.

[9] Moens D. and Vandepitte D., 2005, "A fuzzy finite element procedure for the calculation of uncertain frequency-response functions of damped structures: Part 1—Procedure", Journal of Sound and Vibration, 288, 3, 431-462.

[10] B. Möller, M. Beer, Fuzzy randomness: uncertainty in civil engineering and computational mechanics. Springer Science & Business Media, New York, 2013.

[11] Y. Ben-Haim, Info-Gap Decision Theory 2$^{nd}$ Edition Decisions Under Severe Uncertainty, Academic Press, London, 2006.

[12] C. Soize, Random matrix theory for modeling uncertainties in computational mechanics, Comput. Methods in Appl. Mech. Eng. 194 (12-16) (2005) 1333-1366. https://doi.org/10.1016/j.cma.2004.06.038.

[13] R.S. Langley, V. Cotoni, Response variance prediction for uncertain vibro-acoustic systems using a hybrid deterministic-statistical method, J. Acoust. Soc. Am. 122(6) (2007) 3445-3463. https://doi.org/10.1121/1.2799499.

[14] E. Reynders, J. Legault, R. S. Langley, An efficient probabilistic approach to vibro-acoustic analysis based on the Gaussian orthogonal ensemble, The Journal of the Acoustical Society of America 136, 201 (2014); https://doi.org/10.1121/1.4881930

[15] R. Ghanem, P.D. Spanos, Polynomial Chaos in Stochastic Finite Elements, J. Appl. Mech. 57 (1) (1990) 197-202. https://doi.org/10.1115/1.2888303.

[16] F. P. A. Coolen, On the use of imprecise probabilities in reliability. Qual. Reliab. Eng. Int. 20(193) (2004). 193-202. https://doi.org/10.1002/qre.560.

[17] L.V. Utkin, F.P.A. Coolen. Imprecise reliability: An introductory overview. Computational Intelligence in Reliability Engineering 40 (2007) 261-306. https://doi.org/10.1007/978-3-540-37372-8_10.

[18] E.T. Jaynes, Probability Theory: The Logic of Science, Cambridge University Press, 2003. https://doi.org/10.1017/CBO9780511790423.

[19] A. Cicirello, R. Langley, Vibro-acoustic response of engineering structures with mixed type of probabilistic and nonprobabilistic uncertainty models. ASCE-ASME J. Risk Uncertain. Eng. Syst. B. 1(4) (2015) 13 pages. http://dx.doi.org/10.1115/1.4030470.

[20] W. Dong, H. Shah, Vertex method for computing functions of fuzzy variables, Fuzzy Sets and Syst. 24(1) (1987) 65–78. https://doi.org/10.1016/0165-0114(87)90114-X.

[21] A.I.J. Forrester, A. Sóbester, A.J. Keane, Engineering design via surrogate modelling - A Practical Guide, John Wiley and Sons, University of Southampton (UK), 2008.

[22] D.B. McDonald, W.J. Grantham, W.L. Tabor, Response surface model development for global/local optimization using radial basis functions, AIAA 2000-4747, 2005 https://doi.org/10.2514/6.2000-4776.

[23] H.H. Khodaparast, Y. Govers, I. Dayyani, S. Adhikari, M. Link, M.I. Friswell, J.E. Motthershead, J. Sienz, Fuzzy finite element model updating of the DLR AIRMOD test structure, Appl. Math. Model 52 (2017) 512-526. https://doi.org/10.1016/j.apm.2017.08.001.

[24] Z. Deng, Z. Guo, X. Zhang, Interval model updating using perturbation method and Radial Basis Function neural networks, Mech Syst Signal Process 84 (2017) 699-716. https://doi.org/10.1016/j.ymssp.2016.09.001.

[25] L.Wang, Z. Chen, G. Yang, An interval uncertainty analysis method for structural response bounds using feedforward neural network differentiation, Appl. Math. Model 82 (2020) 449-468. https://doi.org/10.1016/j.apm.2020.01.059

[26] M. De Munck, D. Moens, W. Desmet, D. Vandepitte, An Efficient Response Surface Based Optimisation Method for Non-Deterministic Harmonic and Transient Dynamic Analysis, Comput Model Eng Sci 47(2) (2009) 119-166. https://doi.org/10.3970/cmes.2009.047.119.

[27] M.S. Eldred, L.P. Swiler, G. Tang, Mixed aleatory-epistemic uncertainty quantification with stochastic expansions and optimization-based interval estimation, Reliability Engineering & System Safety, 96(9) (2011) 1092-1113. https://doi.org/10.1016/j.ress.2010.11.010

[28] D.R. Jones, W.J. Welch, M. Schonlau, Efficient global optimization of expensive black-box functions, J. Global Optim. 13 (1998) 455-492. http://dx.doi.org/10.1023/A:1008306431147.

[29] H.P. Wan, Y.Q. Ni, A New Approach for Interval Dynamic Analysis of Train-Bridge System Based on Bayesian Optimization, J. Eng. Mech. 146(5) (2020) 14 pages. https://doi.org/10.1061/(ASCE)EM.1943-7889.0001735.

[30] C.E. Rasmussen, C.K.I Williams, Gaussian Processes for Machine Learning, The MIT Press, Massachusetts Institute of Technology, 2006.

[31] P. Frazier, A Tutorial on Bayesian Optimization (2018), https://arXiv:1807.02811v1

[32] E. Brochu, V.M. Cora, N. de Freitas, A Tutorial on Bayesian Optimization of Expensive Cost Functions, with Application to Active User Modeling and Hierarchical Reinforcement Learning (2010). https://arxiv.org/abs/1012.2599.

[33] H.J. Kushner, A new method of locating the maximum of an arbitrary multipeak curve in presence of noise. J. Basic Engineering 86 (1964) 97-106.

[34] J. Močkus, On bayesian methods for seeking the extremum, in Optimization Techniques IFIP Technical Conference. Springer (1975) 400-404.

[35] P.D. Berger, R.E. Maurer, G.B. Celli, Experimental Design – with applications in management, Engineering, and the Sciences - 2nd Edition, Springer, 2018.

[36] R.N. Raghu, N. Kacker, E.S. Lagergren, J.J. Filliben, Taguchi's orthogonal arrays are classical designs of experiments, J Res Natl Inst Stand Technol 96(5) (1991) 577-591. https://doi.org/10.6028/jres.096.034   .

[37] M. Schonlau, Computer experiments and global optimization, PhD thesis, University of Waterloo (Canada) 1977.

[38] D.D. Cox, S. John, SDO: A statistical method for global optimization. In Multidisciplinary Design Optimization: State of the Art (1997) 315-329.

[39] N. Srinivas, A. Krause, S.M. Kakade, M. Seeger, Gaussian Process Optimization in the Bandit Setting: No regret and Experimental Design, In Proc. International Conference on Machine Learning (ICML), 2010.

[40] O. Zienkiewicz, R. Taylor, J.Z. Zhu, The Finite Element Method: Its Basis and Fundamentals - 7th Edition, Butterworth-Heinemann, 2013.

[41] J. Katsikadelis, The Boundary Element Method for Engineers and Scientists -2nd Edition Theory and Applications, Academic Press, 2016.

[42] B.F. Shorr, The Wave Finite Element Method, Springer, 2004.

[43] K.P. Murphy, Machine Learning A Probabilistic Perspective, 2012.

[44] G. Casella, R.L. Berger, Statistical Inference 2nd Edition (2002)

[45] M. Filippone, M. Zhong, M. Girolami, A comparative evaluation of stochastic-based inference methods for Gaussian process models, Mach Learn 93 (2013) 93-114. https://doi.org/10.1007/s10994-013-5388-x.

[46] I. Murray, R.P. Adams, Slice sampling covariance hyperparameters of latent Gaussian models, https://arxiv.org/abs/1006.0868, 2010.

[47] Z. Wang, M. Zoghi, D. Matheson, N. de Freitas, Bayesian Optimization in High Dimensions via Random Embeddings, IJCAI-13: proceedings of the Twenty-Third International Joint Conference on Artificial Intelligence: Beijing, China, 3-9 August 2013. - Vol. 3.

[48] D.R. Burt, C.E. Rasmussen, M. Van der Wilk, Rates of Convergence for Sparse Variational Gaussian Process Regression, https://arxiv.org/abs/1903.03571, 2019.

[49] V. Pincheny, T. Wagner, D. Ginsbourger, A benchmark of kriging-based infill criteria for noisy optimization, Struct. Multidiscip. O. 48 (2013) 607-626. https://doi.org/10.1007/s00158-013-0919-4.

[50] B. Shahriari, K. Swersky, Z. Wang, R. P. Adams and N. de Freitas, "Taking the Human Out of the Loop: A Review of Bayesian Optimization," in Proceedings of the IEEE, vol. 104, no. 1, pp. 148-175, Jan. 2016, doi: 10.1109/JPROC.2015.2494218.

[51] H. Wang, B. van Stein, M. Emmerich, T. Bäck, A new acquisition function for Bayesian optimization based on the moment-generating function, 2017 IEEE International Conference on Systems, Man, and Cybernetics (SMC) (2017) 507-512. doi: https://10.1109/SMC.2017.8122656.

[52] R.J. Jones. A Taxonomy of global optimization methods based on response surfaces, J. Global Optim. 21 (2001) 345-383. https://doi.org/10.1023/A:1012771025575.

[53] A. D. Bull, Convergence rates of efficient global optimization algorithms. Journal of Machine Learning Research 12 (2011) 2879-2904. https://arxiv.org/abs/1101.3501.

[54] M.J. Sasena, P. Papalambrod, P. Goovaerts, Exploration of Metamodeling Sampling Criteria for Constrained Global Optimization. Engineering Optimization 34(3) (2002) 263-278.

[55] Carl Q. Howard, Benjamin S. Cazzolato, Acoustic Analyses Using Matlab® and Ansys®, CRC Press, 2014.

[56] MATLAB, 2019b, The MathWorks Inc., Natick, Massachusetts, 2019.

[57] D.G. Inman. Engineering Vibrations, 4th edition, London: Pearson Education, 2014.

[58] W.L. Brogan. Modern Control Theory, 3rd edition. New Jersey: Prentice-Hall, 1991.

[59] D.R. Bland. The theory of linear viscoelasticity. Pergamon press, 1960.

# Appendix – Taguchi matrices for plate-acoustic-cavity system

The $\mathbf{B}_0$ selected by using the Taguchi matrices $\mathbf{L}_s(q^r)$ in the second numerical application are reported below, with $r = 4$; and $q$ varied from 2 to 5. The minimum $s = 1 + r(q-1)$, and ultimately $s$ chosen as a multiple of $q$.

| $\mathbf{B}_0$ for D8 $\mathbf{L}_8(2^4)$ | mm | Pa | kg/m^3 | m/s |
|---|---|---|---|---|
| 1 | $280 \times 10^{-2}$ | $700.000 \times 10^8$ | $120.000 \times 10^{-2}$ | 342.0 |
| 2 | $280 \times 10^{-2}$ | $700.000 \times 10^8$ | $120.000 \times 10^{-2}$ | 346.0 |
| 3 | $280 \times 10^{-2}$ | $719.000 \times 10^8$ | $122.000 \times 10^{-2}$ | 342.0 |
| 4 | $280 \times 10^{-2}$ | $719.000 \times 10^8$ | $122.000 \times 10^{-2}$ | 346.0 |
| 5 | $320 \times 10^{-2}$ | $700.000 \times 10^8$ | $122.000 \times 10^{-2}$ | 342.0 |
| 6 | $320 \times 10^{-2}$ | $700.000 \times 10^8$ | $122.000 \times 10^{-2}$ | 346.0 |
| 7 | $320 \times 10^{-2}$ | $719.000 \times 10^8$ | $120.000 \times 10^{-2}$ | 342.0 |
| 8 | $320 \times 10^{-2}$ | $719.000 \times 10^8$ | $120.000 \times 10^{-2}$ | 346.0 |

| $\mathbf{B}_0$ for D9 $\mathbf{L}_9(3^4)$ | mm | Pa | kg/m^3 | m/s |
|---|---|---|---|---|
| 1 | $280 \times 10^{-2}$ | $700.000 \times 10^8$ | $122.000 \times 10^{-2}$ | 342.0 |
| 2 | $280 \times 10^{-2}$ | $709.500 \times 10^8$ | $121.000 \times 10^{-2}$ | 344.0 |
| 3 | $280 \times 10^{-2}$ | $719.000 \times 10^8$ | $120.000 \times 10^{-2}$ | 346.0 |
| 4 | $300 \times 10^{-2}$ | $700.000 \times 10^8$ | $121.000 \times 10^{-2}$ | 346.0 |
| 5 | $300 \times 10^{-2}$ | $709.500 \times 10^8$ | $122.000 \times 10^{-2}$ | 342.0 |
| 6 | $300 \times 10^{-2}$ | $719.000 \times 10^8$ | $120.000 \times 10^{-2}$ | 344.0 |
| 7 | $320 \times 10^{-2}$ | $700.000 \times 10^8$ | $122.000 \times 10^{-2}$ | 344.0 |
| 8 | $320 \times 10^{-2}$ | $709.500 \times 10^8$ | $120.000 \times 10^{-2}$ | 346.0 |
| 9 | $320 \times 10^{-2}$ | $719.000 \times 10^8$ | $121.000 \times 10^{-2}$ | 342.0 |

| $B_0$ for D16 $L_{16}(4^4)$ | mm | Pa | kg/m^3 | m/s |
|---|---|---|---|---|
| 1 | $280\times10^{-2}$ | $700.000\times10^8$ | $120.000\times10^{-2}$ | 342.0 |
| 2 | $280\times10^{-2}$ | $706.330\times10^8$ | $120.670\times10^{-2}$ | 343.3 |
| 3 | $280\times10^{-2}$ | $712.670\times10^8$ | $121.330\times10^{-2}$ | 344.7 |
| 4 | $280\times10^{-2}$ | $719.000\times10^8$ | $122.000\times10^{-2}$ | 346.0 |
| 5 | $290\times10^{-2}$ | $700.000\times10^8$ | $120.670\times10^{-2}$ | 344.7 |
| 6 | $290\times10^{-2}$ | $706.330\times10^8$ | $121.330\times10^{-2}$ | 346.0 |
| 7 | $290\times10^{-2}$ | $712.670\times10^8$ | $122.000\times10^{-2}$ | 342.0 |
| 8 | $290\times10^{-2}$ | $719.000\times10^8$ | $120.000\times10^{-2}$ | 343.3 |
| 9 | $310\times10^{-2}$ | $700.000\times10^8$ | $121.330\times10^{-2}$ | 342.0 |
| 10 | $310\times10^{-2}$ | $706.330\times10^8$ | $122.000\times10^{-2}$ | 343.3 |
| 11 | $310\times10^{-2}$ | $712.670\times10^8$ | $120.000\times10^{-2}$ | 344.7 |
| 12 | $310\times10^{-2}$ | $719.000\times10^8$ | $120.670\times10^{-2}$ | 346.0 |
| 13 | $320\times10^{-2}$ | $700.000\times10^8$ | $122.000\times10^{-2}$ | 344.7 |
| 14 | $320\times10^{-2}$ | $706.330\times10^8$ | $120.000\times10^{-2}$ | 346.0 |
| 15 | $320\times10^{-2}$ | $712.670\times10^8$ | $120.670\times10^{-2}$ | 342.0 |
| 16 | $320\times10^{-2}$ | $719.000\times10^8$ | $121.330\times10^{-2}$ | 343.3 |

| $B_0$ for D25 $L_{25}(5^4)$ | mm | Pa | kg/m^3 | m/s |
|---|---|---|---|---|
| 1 | $280\times10^{-2}$ | $700.000\times10^8$ | $120.000\times10^{-2}$ | 342.0 |
| 2 | $280\times10^{-2}$ | $704.750\times10^8$ | $120.500\times10^{-2}$ | 343.0 |
| 3 | $280\times10^{-2}$ | $709.500\times10^8$ | $121.000\times10^{-2}$ | 344.0 |
| 4 | $280\times10^{-2}$ | $714.250\times10^8$ | $121.500\times10^{-2}$ | 345.0 |
| 5 | $280\times10^{-2}$ | $719.000\times10^8$ | $122.000\times10^{-2}$ | 346.0 |
| 6 | $290\times10^{-2}$ | $700.000\times10^8$ | $120.500\times10^{-2}$ | 344.0 |
| 7 | $290\times10^{-2}$ | $704.750\times10^8$ | $121.000\times10^{-2}$ | 345.0 |
| 8 | $290\times10^{-2}$ | $709.500\times10^8$ | $121.500\times10^{-2}$ | 346.0 |
| 9 | $290\times10^{-2}$ | $714.250\times10^8$ | $122.000\times10^{-2}$ | 342.0 |
| 10 | $290\times10^{-2}$ | $719.000\times10^8$ | $120.000\times10^{-2}$ | 343.0 |
| 11 | $300\times10^{-2}$ | $700.000\times10^8$ | $121.000\times10^{-2}$ | 346.0 |
| 12 | $300\times10^{-2}$ | $704.750\times10^8$ | $121.500\times10^{-2}$ | 342.0 |
| 13 | $300\times10^{-2}$ | $709.500\times10^8$ | $122.000\times10^{-2}$ | 343.0 |
| 14 | $300\times10^{-2}$ | $714.250\times10^8$ | $120.000\times10^{-2}$ | 344.0 |
| 15 | $300\times10^{-2}$ | $719.000\times10^8$ | $120.500\times10^{-2}$ | 345.0 |
| 16 | $310\times10^{-2}$ | $700.000\times10^8$ | $121.500\times10^{-2}$ | 343.0 |
| 17 | $310\times10^{-2}$ | $704.750\times10^8$ | $122.000\times10^{-2}$ | 344.0 |
| 18 | $310\times10^{-2}$ | $709.500\times10^8$ | $120.000\times10^{-2}$ | 345.0 |
| 19 | $310\times10^{-2}$ | $714.250\times10^8$ | $120.500\times10^{-2}$ | 346.0 |
| 20 | $310\times10^{-2}$ | $719.000\times10^8$ | $121.000\times10^{-2}$ | 342.0 |
| 21 | $320\times10^{-2}$ | $700.000\times10^8$ | $122.000\times10^{-2}$ | 345.0 |
| 22 | $320\times10^{-2}$ | $704.750\times10^8$ | $120.000\times10^{-2}$ | 346.0 |
| 23 | $320\times10^{-2}$ | $709.500\times10^8$ | $120.500\times10^{-2}$ | 342.0 |
| 24 | $320\times10^{-2}$ | $714.250\times10^8$ | $121.000\times10^{-2}$ | 343.0 |
| 25 | $320\times10^{-2}$ | $719.000\times10^8$ | $121.500\times10^{-2}$ | 344.0 |